\long\def\symbolfootnote[#1]#2{\begingroup%
\def\thefootnote{\fnsymbol{footnote}}\footnote[#1]{#2}\endgroup} 

\def\henon{H\'{e}non}
\def\Eg{E_\gamma}
\def\dr{{\rm d}}
\def\ra{r_{\rm a}}
\def\rh{r_{\rm h}}
\def\rj{r_{\rm J}}
\def\rt{r_{\rm t}}
\def\rv{r_{\rm v}}
\def\trh{\tau_{\rm rh}}
\def\rvec{\textbf{\emph{r}}}
\def\vvec{\textbf{\emph{v}}}
\def\xvec{\textbf{\emph{x}}}

\documentclass[useAMS,usenatbib, times]{mn2e}
\usepackage{amssymb,latexsym,graphicx,natbib,eufrak,times,amsmath,xfrac,nicefrac}
\usepackage[caption=false]{subfig}
\usepackage{url}
\usepackage{array}
\usepackage{siunitx}

\title[Testing lowered isothermal models with direct N-body simulations of globular clusters]{Testing lowered isothermal models with direct N-body simulations of globular clusters}
\author[Alice Zocchi, Mark Gieles, Vincent H\'enault-Brunet, Anna Lisa Varri]
  {Alice Zocchi$^1$\thanks{E-mail:
a.zocchi@surrey.ac.uk},  Mark Gieles$^1$, Vincent H\'enault-Brunet$^1$,  Anna Lisa Varri$^2$ \\
$^1$ Department of Physics, University of Surrey, Guildford, GU2 7XH,UK.\\
$^2$ School of Mathematics and Maxwell Institute of Mathematical Sciences, University of Edinburgh, King’s Buildings, Edinburgh EH9 3JZ, UK.\\
}
\date{Accepted 2016 May 6. Received 2016 May 6; in original form 2016 March 14}

\def\LaTeX{L\kern-.36em\raise.3ex\hbox{a}\kern-.15em
    T\kern-.1667em\lower.7ex\hbox{E}\kern-.125emX}

\begin{document}         
\maketitle
\begin{abstract}
Several self-consistent models have been proposed, aiming at describing the phase space distribution of stars in globular clusters. This study explores the ability of the recently proposed \textsc{limepy} models (Gieles \& Zocchi) to reproduce the dynamical properties of direct $N$-body models of a cluster in a tidal field, during its entire evolution. These  dynamical models include prescriptions for the truncation and the degree of radially-biased anisotropy contained in the system, allowing us to explore the interplay between the role of anisotropy and tides in various stages of the life of star clusters. We show that the amount of anisotropy in an initially tidally underfilling cluster increases in the pre-collapse phase, and then decreases with time, due to the effect of the external tidal field on its spatial truncation. This is reflected in the correspondent model parameters, and the best-fit models reproduce the main properties of the cluster at all stages of its evolution, except for the phases immediately preceding and following core collapse. We also notice that the best-fit \textsc{limepy} models are significantly different from isotropic King models, especially in the first part of the evolution of the cluster. Our results put limits on the amount of radial anisotropy that can be expected for clusters evolving in a tidal field, which is important to understand other factors that could give rise to similar observational signatures, such as the presence of an intermediate-mass black hole.
\end{abstract}
\begin{keywords}
galaxies: star clusters -- globular clusters: general  -- methods: numerical -- stars: kinematics and dynamics
\end{keywords}

\section{Introduction}
\label{Sect_Introduction}

For a long time, globular clusters have been considered to be simple, spherical, and isotropic systems. However, it is well known that several factors contribute to their evolution, causing their properties to be far from this simple approximation. In particular, the effects of the external tidal field, of pressure anisotropy, and of rotation are particularly important in shaping their kinematical properties. In the present work, we focus on the first two ingredients, and we explore their role during the entire life of globular clusters.

The presence of radially-biased pressure anisotropy in globular clusters may be interpreted both as a signature of their formation through violent relaxation \citep{Lynden-Bell1967} and as a product of their dynamical evolution. The role of violent relaxation in globular clusters has been explored by a number of numerical studies \cite[for example, see][among many others]{Aarseth1988,Vesperini1996,McMillan2007}. Recently, by analysing the results of several numerical simulations, \cite{Vesperini2014} have studied the dynamics of a cluster in the phase of violent relaxation, and showed that radially-biased anisotropy naturally arises during this process. They point out that, after having experienced violent relaxation, isolated systems are characterised by isotropy in the centre and increasing radially-biased anisotropy in the outer parts \citep[see also][]{VanAlbada1982,Trenti2005}; for systems that evolve in a tidal field, anisotropy reaches a maximum and then decreases again, with the outermost regions being isotropic. 

Pioneering numerical investigations have shown that pressure anisotropy becomes important during the evolution of the clusters, even when the initial conditions are spherical and isotropic \citep{Henon1971a,SpitzerHart1971,SpitzerShull1975}. \cite{Spitzer1987} showed that during their evolution, isolated globular clusters develop a structure composed by two distinct regions: an isotropic core, and a radially anisotropic halo of stars. An explanation for this is that stars are scattered from the centre in the halo on radial orbits, as a consequence of the gravothermal instability which is thought to be responsible for the core collapse \citep{Lynden-Bell1967,Lynden-BellWood1968,SpitzerShapiro1972}. Simulations taking into account stars with different masses by \cite{GierszHeggie1996} confirmed this picture, showing that the post-collapse evolution and anisotropy profile are self-similar, with the system being more anisotropic at larger distances from the centre. 

Simulations taking into account the presence of an external tidal field have unveiled two possible scenarios for the development of pressure anisotropy, depending on the strength of the tidal field. On the one hand, \cite{GierszHeggie1997} and \cite{TakahashiLeeInagaki1997} showed that during the collapse phase and shortly after the cluster is isotropic in the centre, and increasingly radially anisotropic at larger radii. As the evolution proceeds, the external tidal field has the effect of suppressing the anisotropy, and the system eventually becomes fully isotropic. This happens for two main reasons: first, mass loss has the effect of exposing the deeper parts of the systems, where deviations from isotropy are more modest \citep{GierszHeggie1997}; second, the tidal torque induces isotropy in the velocity dispersion of the outer regions of the cluster, as described by \cite{OhLin1992} and \cite{Pontzen2015}. On the other hand, other sets of numerical simulations \citep[for example][and many others]{Takahashi1997,AarsethHeggie1998,TakahashiLee2000,BaumgardtMakino2003} have shown that during their evolution clusters remain basically isotropic everywhere, except for the outermost parts, where tangentially biased anisotropy is present, due to the fact that stars on radial orbits are preferentially lost as effect of the interaction with the tidal field. Recently, \cite{Sollima2015} presented two simulations in which different flavours of anisotropy develop, pointing out that the type of anisotropy arising during the evolution of the system is related to the strength of the tidal field acting on it. As also shown by \cite{Tiongco2016}, clusters that are originally tidally underfilling develop a significant amount of radial anisotropy, while those that are initially filling their Roche volume remain basically isotropic throughout their evolution.

Recently, it has become feasible to measure anisotropy in the very central regions of nearby clusters \citep[i.e., typically within $100\arcsec$ from the centre, see][]{Watkins2015}, with the Hubble Space Telescope (HST). In the near future, data from the Global Astrometric Interferometer for Astrophysics (Gaia), will enable us to obtain such measurements in a large number of clusters \citep{PancinoGAIA2013,Sollima2015}, and for stars located especially in their outer regions, where the effects described above can be observed and discriminated. 

Given the complexity of this picture, and to be able to analyse the data in the best possible way, it is crucial to have at our disposal adequate models to describe the different states in the dynamical evolution of globular clusters, in order to give an accurate representation of their observed properties. The instantaneous properties of these stellar systems are well described by relatively simple distribution function based models, defined with few assumptions. Spherical, isotropic, and non-rotating King models \citep{Michie1963,King1966} have been widely used to describe Galactic globular clusters \citep[see for example][]{MLvdM2005,ZBV2012}. Non-rotating Wilson models, having a different truncation prescription with respect to King models, have also proven to be adequate to reproduce especially the outermost slope of the surface brightness profiles of some clusters \citep{MLvdM2005}. In some cases, to explain the observed kinematic properties of clusters \citep{Sollima2009,ZBV2012,Ibata2013,BVBZ2013}, it is necessary to consider anisotropic models, such as those introduced by \cite{Michie1963} and \cite{BertinTrenti2003}, or rotating models, such as those by \cite{Wilson1975} and \cite{VB2012}. Moreover, when the effects of mass segregation cannot be neglected, multi-mass models \citep{DaCostaFreeman1976,GunnGriffin1979} are necessary to properly describe Galactic globular clusters \citep{PryorMeylan1993,Sollima2012,Sollima2015}. 

In this study we will perform a comparative investigation between a family of distribution function based dynamical models, recently proposed by \cite{GielesZocchi2015}, and an $N$-body simulation performed by \cite{AlexanderGieles2012}, which offers a representation of the entire dynamical evolution of an idealised star cluster. Such a family of dynamical models is characterised by a parametrised truncation prescription, and a variable degree of radially-biased pressure anisotropy. Their flexibility makes these models particularly suitable for studying the evolution of the main structural and kinematical properties of a star cluster in the presence of a mild tidal perturbation. In principle, these models allow also for the presence of multiple mass components, but, in the present study, we will consider only the simple single-component case, as appropriate for the comparison with an equal-mass $N$-body model. 

Our purpose is to assess whether these models may be used to describe collisional stellar systems in different dynamical states, and thus to define a parametric evolutionary sequence of well-posed dynamical equilibria, which may also be used as a tool to characterise the entire dynamical evolution of a star cluster in a simplified, yet physically motivated, way. This idea has been already explored, in the past, starting with \cite{Prata1971}, who proposed a method to calculate the dynamical evolution of star clusters by means of a sequence of King models. \cite{Chernoff1986} studied the pre-collapse phase of evolution of clusters and the influence of tidal heating on their relaxation, and later \cite{Chernoff1987} traced the evolution of clusters to core collapse and tidal disruption by means of a three-parameter series of King models. \cite{Wiyanto1985} compared the evolution of King models by using the isotropised orbit-averaged Fokker-Planck equation, and \cite{Vesperini1997} proposed a comparison to the results of numerical simulations with a broad range of initial conditions. In this framework, the present work expands this investigation by introducing two additional elements of interest: the presence of anisotropy and a certain degree of freedom in the definition of the truncation prescription. Moreover, with respect to the previous studies, we extend our analysis to the entire life of a star cluster, by considering also the post-collapse evolution until final dissolution in the tidal field.

The paper is organised as follows: in Section~\ref{Sect_Models} we introduce the models, and in Section~\ref{Sect_Simulations} we illustrate the main properties of the simulations we consider in our analysis. In Section~\ref{Sect_FittingMethods} we describe the fitting method we adopted, and in Section~\ref{Sect_Results} we show and discuss our results. Our conclusions are presented in Section~\ref{Sect_Conclusion}. Finally, Appendix~\ref{App_residuals} provides a more quantitative comparison between relevant quantities for the models and for the simulation.


\section{Dynamical models}
\label{Sect_Models}

We consider the family of spherical dynamical models presented by \cite{GielesZocchi2015}. In the following, we will refer to these models as \textsc{limepy} models, from the name of the code that is used to calculate them\footnote{The \textsc{limepy} (Lowered Isothermal Model Explorer in PYthon) code is available from \url{https://github.com/mgieles/limepy}}. The distribution function, depending on the specific energy $E$ and angular momentum $J^2$, is
\begin{equation}
f(E,J^2) =  A \, \exp \left(-\frac{J^2}{2 \ra^2 s^2}\right) \Eg \left(g, \dfrac{\phi(\rt)-E}{s^2}\right) \ ,
\label{Eq_DF_Limepy}
\end{equation}
for $E<\phi(\rt)$, with $\rt$ the truncation radius, and $0$ otherwise. We recall that the energy can be expressed as $E = v^2/2 + \phi(r)$, where $\phi$ is the specific potential and $v$ is the velocity, and the angular momentum as $J^2 = r^2 v^2_{\rm t}$, where we used the tangential component of the velocity $v_{\rm t}$, and the radius $r$, indicating the distance from the centre of the system. In equation~(\ref{Eq_DF_Limepy}) we have introduced the function 
\begin{align}
\Eg(g, x) =  
\begin{cases}
\exp(x)  &  g=0 \\
\displaystyle\exp(x) \frac{\gamma(g, x)}{\Gamma(g)}   &  g>0,
\end{cases}
\label{eq:eg}
\end{align}
where $\gamma(g, x)$ is the lower incomplete gamma function, and $\Gamma(g)$ is the gamma function \citep[for properties of these functions, see][]{EgammaModels,AbramowitzStegun}.

To identify one model within the family, it is necessary to specify the values of three parameters: 
\begin{itemize}
 \item $W_0$ is the central dimensionless potential, and sometimes it is referred to as the \textit{concentration parameter} of the model. This parameter is used as a boundary condition to solve the Poisson equation, and it determines the shape of the radial profiles of some relevant quantities; 
 \item the \textit{anisotropy radius} $\ra$ is related to the amount of anisotropy present in the system. The smaller it is, the more anisotropic is the model. In the limit $\ra \rightarrow \infty$, the models become isotropic; in practice, a configuration may be safely considered isotropic if the anisotropy radius is larger than the truncation radius $\rt$;
 \item the \textit{truncation parameter} $g$ sets the sharpness of the truncation in energy: the larger it is, the more extended the models are, and the less abrupt the truncation is. When considering the isotropic version of the models, $g=0$ corresponds to the \citet{Woolley1954} models, $g=1$ to the \citet{King1966} models, and $g=2$ to the non-rotating \citet{Wilson1975} models. Depending on the value of $W_0$, there exists a maximum value of $g$ that is allowed to consider when finite models are needed: for $g\lesssim2.1$ the models are finite for all the values of $W_0$, but for low values of $W_0$ larger values of $g$ are also acceptable \citep[see fig.~4 of][]{GielesZocchi2015}.
\end{itemize}
It is important to note that a model with given values of $W_0$ and $g$ becomes more extended when the value of $\ra$ is decreased. The same effect is obtained when $g$ is increased, keeping $W_0$ and $\ra$ fixed. This causes a degeneracy in the parameter space when fitting models only to surface brightness or number density profiles: without information on the three-dimensional kinematics of a system, it is impossible to disentangle the two effects \citep[for a discussion, see also][]{ZBV2012}.

In addition, it is necessary to specify two scales, that are related to $A$ and $s$, which represent a mass density in the phase space and a velocity scale; in particular, for models with high concentration, $s$ is approximately equal to the one-dimensional velocity dispersion in the centre of the system. These scales naturally define also a radial scale. In this way, the velocity, radial, and mass units are defined, so that every property of the model can be expressed in these terms. 
We note that these three physical units are not independent, and that two of them fully specify the free scales of the models \citep[for details, see][]{GielesZocchi2015}. In fitting these models to data, it is useful to consider the mass scale and the radial scale as fitting parameters, so we will consider these as the physical scales in the remainder of this paper.

Starting from the distribution function of equation~(\ref{Eq_DF_Limepy}), we can calculate several quantities that are useful to describe a stellar system, and that are normally used when comparing models to observations. In particular, we consider in the following the mass density $\rho$ and the radial and tangential components of the velocity dispersion $\sigma_{\rm r}$ and $\sigma_{\rm t}$ \citep[for a definition, see][]{GielesZocchi2015,BT1987}; we will also use $\sigma$, defined from the two components of the velocity dispersion ($\sigma^2 = \sigma_{\rm r}^2 + \sigma_{\rm t}^2$). 

These models are isotropic in the centre, radially anisotropic in the intermediate part, and isotropic again near the truncation radius\footnote{The shape of this profile is a consequence of the definition of the distribution function. Near the truncation radius, the models behave like polytropes and are isotropic \citep[for a more detailed explanation, see also Sect.~2.1.5 of][]{GielesZocchi2015}.} $\rt$. This property is particularly interesting when using them to describe globular clusters because, as outlined in Section~\ref{Sect_Introduction}, radial anisotropy is expected to develop in the intermediate parts of these systems during the early stages of their evolution, while the innermost and outermost parts are expected to be isotropic because of relaxation processes and tidal effects, respectively. The degree of anisotropy of the configurations may be characterised, as usual, by means of a local or a global diagnostics. In the present analysis, we will adopt the following definition of the anisotropy parameter:
\begin{equation}
 \beta = 1 - \frac{\sigma^2_{\rm t}}{2 \sigma^2_{\rm r}} \ ,
\label{Eq_beta}
\end{equation}
such that  a positive value of $\beta$ indicates radial anisotropy, a negative value tangential anisotropy, and $\beta = 0$ isotropy. In addition, we will also adopt the global quantity introduced by \citet{PolyachenkoShukhman1981}
\begin{equation}
 \kappa = \frac{2 K_{\rm r}}{K_{\rm t}} \ ,
 \label{Eq_kappa}
\end{equation}
where $K_{\rm r}$ and $K_{\rm t}$ are the radial and tangential components of the kinetic energy, respectively. Isotropy is characterised by $\kappa = 1$, radial anisotropy by $\kappa > 1$ and tangential anisotropy by $\kappa < 1$. An anisotropic model is stable against radial orbit instability if it satisfies the criterion introduced by \citet{PolyachenkoShukhman1981}, $\kappa < 1.7 \pm 0.25$: all the models considered in this paper satisfy this criterion and are therefore stable against radial orbit instability.

Two of the main simplifications of these models are the assumptions of spherical symmetry and the absence of rotation. For real globular clusters it may be important to relax these assumptions, in order to give a realistic representation of their structure and dynamics. For the purpose of this study, however, we choose to rely on these assumptions, especially in consideration of properties of the reference $N$-body model we wish to analyse (which is spherical and non-rotating). The focus of our investigation is indeed the role of pressure anisotropy, and \textsc{limepy} models, especially by virtue of their parameters $g$ and $\ra$, are the ideal framework to quantify the deviations from isotropy in velocity space emerging during the life of a star cluster, and to describe their evolution in the presence of an external tidal field.


\section{Numerical simulations}
\label{Sect_Simulations}
We consider the results of a simulation published in \citet{AlexanderGieles2012}, and run with Aarseth's \textsc{nbody6} \citep{MakinoAarseth1992, Aarseth1999, Aarseth2003}. The starting configuration of the simulation is a cluster composed of $N = 65\,536$ stars with the same mass, distributed according to a \cite{Plummer1911} spherical model. The simulation does not include primordial binaries, nor a central black hole. The system is assumed to be on a circular orbit in a weak tidal field generated by a point-mass galaxy; initially, the ratio of the Jacobi radius to the half-mass radius for the cluster is set to $\rj/\rh = 100$. 

The units used in this paper are the conventional \henon\ $N$-body units: $G = M = -4 \, \mathcal{E} = 1$ \citep{Henon1971}, where $G$, $M$, and $\mathcal{E}$ denote the gravitational constant, the total initial mass, and the total energy, respectively. The equations of motion are solved in a reference frame that co-rotates with the orbit of the cluster. The model is initially in virial equilibrium, such that the virial radius $\rv=-GM^2/(2U)=1$, where $U$ is the gravitational energy. Stars are removed from the simulation once they reach $r > 2 \, \rj$ and the simulation is run until complete dissolution, which is defined as $N<100$, and occurs at approximately $t = 6 \times 10^6$ $N$-body times.

The evolution of the cluster is driven by two-body relaxation, three-body binaries, and the interaction with the tidal field. The effects of stellar evolution are not taken into account. After undergoing core collapse, roughly at $t = 1.25 \times 10^4$ $N$-body times, the system expands until it fills its Roche-volume. Although Roche-filling is not clearly defined, at $t \sim 3\times10^5$ $N$-body times $\rh \gtrsim 0.13 \, \rj$, from which moment the ratio $\rh/\rj$ remains roughly constant. 

Here, we consider 21 snapshots of this evolving system, with the aim of determining their dynamical properties. We select ten snapshots before core collapse, equally spaced in time with intervals of $10^3$ $N$-body times, and eleven snapshots after core collapse, with a time spacing of $5 \times 10^4$ $N$-body times. This choice is motivated by the need of accurately sampling the entire life of the cluster with a limited number of snapshots. The first snapshot we consider is at $10^3$ $N$-body times, and the last one is at $5.5 \times 10^5$ $N$-body times, just before the complete dissolution of the system\footnote{All snapshots are available from the Gaia Challenge Wiki page: \url{http://astrowiki.ph.surrey.ac.uk/dokuwiki}}. For each  particle, the complete set of coordinates in phase space are available: the three spatial coordinates $(x, y, z)$, and the three velocity coordinates $(v_x, v_y, v_z)$. Table~\ref{Tab_True_Properties} lists some properties of these snapshots. 

\begin{table}
\begin{center}
\caption[Properties of the considered snapshots.]{Properties of the considered snapshots. For each snapshot, identified by a label in the first column, we list the time at which it was taken, $t$, the total mass of the stars that are bound to the cluster $M$, the half-mass radius $\rh$, the Jacobi radius $\rj$, and the logarithm of the relaxation time at the half-mass radius at that moment, $\trh$. All the quantities are expressed in \henon\ $N$-body units; the horizontal line separates snapshots taken before (also indicated with the string ``pre-CC'') and after core collapse.} 
\label{Tab_True_Properties}
\begin{tabular}{cccccc}
\hline\hline
Snapshot  & $t$ & $M$ & $\rh$ & $\rj$  &  $\log \trh$ \\
\hline
 pre-CC 1 & 1 $\times10^3$ & 1.000 & 0.775 & 77.992 & 2.842 \\
 pre-CC 2 & 2 $\times10^3$ & 0.999 & 0.770 & 77.983 & 2.838 \\
 pre-CC 3 & 3 $\times10^3$ & 0.999 & 0.767 & 77.966 & 2.835 \\
 pre-CC 4 & 4 $\times10^3$ & 0.997 & 0.771 & 77.935 & 2.838 \\
 pre-CC 5 & 5 $\times10^3$ & 0.996 & 0.779 & 77.889 & 2.845 \\
 pre-CC 6 & 6 $\times10^3$ & 0.993 & 0.783 & 77.823 & 2.847 \\
 pre-CC 7 & 7 $\times10^3$ & 0.989 & 0.791 & 77.718 & 2.854 \\
 pre-CC 8 & 8 $\times10^3$ & 0.985 & 0.818 & 77.614 & 2.875 \\
 pre-CC 9 & 9 $\times10^3$ & 0.979 & 0.843 & 77.456 & 2.893 \\
pre-CC 10 & 1 $\times10^4$ & 0.973 & 0.875 & 77.280 & 2.916 \\
\hline
 1 & 5   $\times10^4$ & 0.712 & 2.87 & 70.793 & 3.638 \\
 2 & 1   $\times10^5$ & 0.565 & 3.98 & 65.538 & 3.813 \\
 3 & 1.5 $\times10^5$ & 0.461 & 5.09 & 61.394 & 3.940 \\
 4 & 2   $\times10^5$ & 0.383 & 5.58 & 57.786 & 3.970 \\
 5 & 2.5 $\times10^5$ & 0.314 & 5.97 & 54.215 & 3.981 \\
 6 & 3   $\times10^5$ & 0.248 & 6.18 & 50.472 & 3.966 \\
 7 & 3.5 $\times10^5$ & 0.189 & 6.07 & 46.263 & 3.911 \\
 8 & 4   $\times10^5$ & 0.139 & 5.22 & 41.896 & 3.764 \\
 9 & 4.5 $\times10^5$ & 0.089 & 4.96 & 36.393 & 3.664 \\
10 & 5   $\times10^5$ & 0.048 & 4.01 & 29.958 & 3.436 \\
11 & 5.5 $\times10^5$ & 0.015 & 2.77 & 20.712 & 3.037 \\
\hline
\end{tabular}
\end{center}
\end{table}

As mentioned before, stars are removed from the simulation when they reach a distance of $2 \, \rj$ from the centre of the cluster: this means that every snapshot contains stars that are located outside the Jacobi radius and that are not bound to the system. Moreover, during the evolution of the cluster, it is possible to identify a population of stars having energy in excess of the escape energy of the cluster, $E_{\rm crit} = - (3/2) GM/\rj$: the escape time for these stars can be very long, and therefore they are not instantly ejected \cite[these stars are also called potential escapers, e.g., see][]{FukushigeHeggie2000}. The dynamical models introduced above are designed to describe a system of bound stars, and therefore we consider in our analysis only the stars that have energy below the critical energy for escape. A discussion on the dynamical properties of potential escapers, and on the effects they have on the observable properties of clusters will be presented in a separate paper \citep[Claydon et al., in preparation; see also][]{Kuepper2010}. Moreover, in order to simplify the analysis, we decided to neglect the binaries that formed in the cluster. This choice does not have an impact on our results, because the number of binaries is always very small with respect to the total number of stars in the cluster (with 15 binaries, Snapshot 1 is the one containing the largest number of binaries).

\section{Fitting methods}
\label{Sect_FittingMethods}

\begin{figure*}
\includegraphics[width=0.85\textwidth]{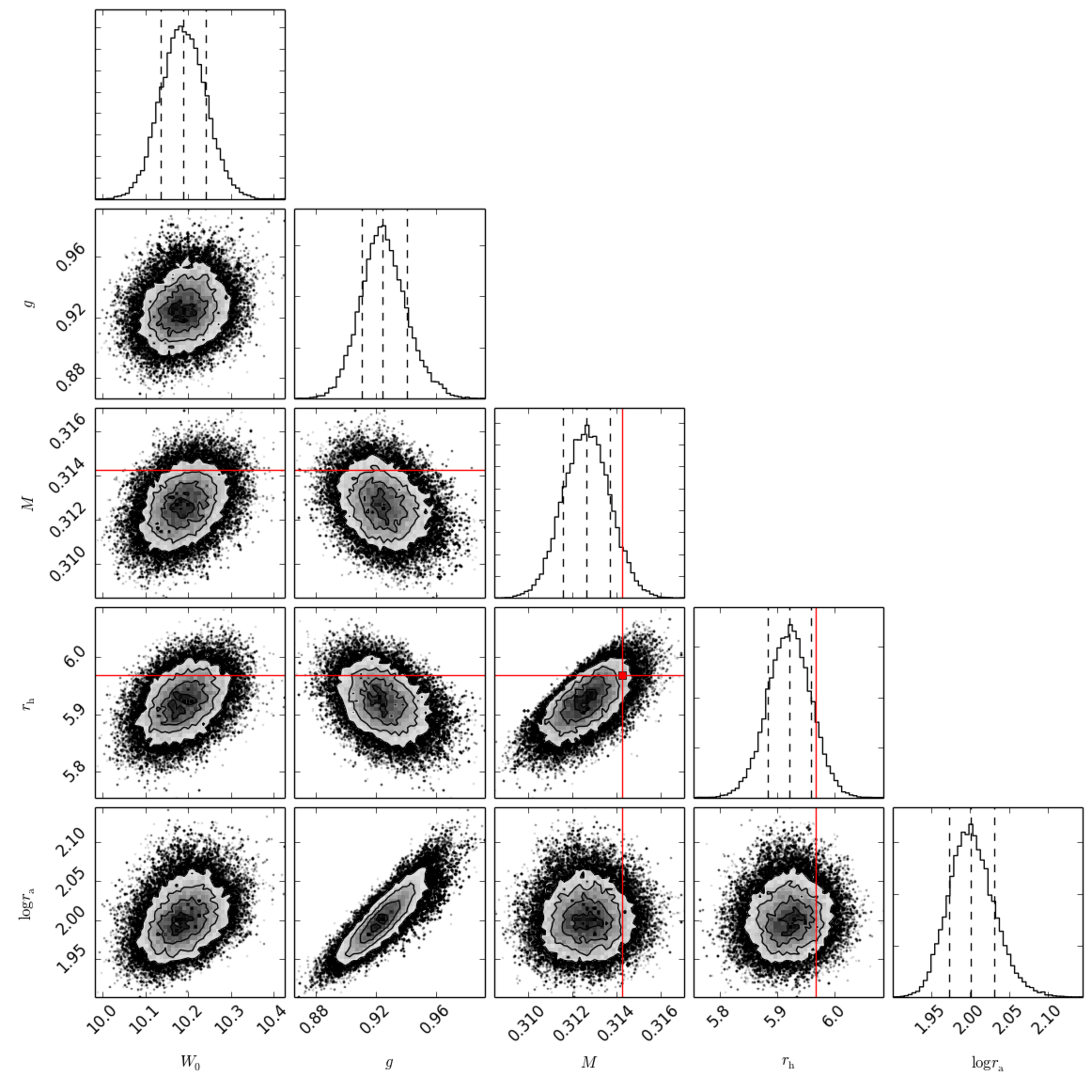}
\caption{Projections of the posterior probability distribution on the planes determined by every pair of parameters (see Section~\ref{Sect_FittingMethods} for the list of fitting parameters, and Section~\ref{Sect_Models} for their description). The plots show the results of the fit carried out with \textsc{emcee} on snapshot number 5 after core collapse, at $t \sim 2.5\times10^5$ $N$-body times. Contours are shown at 0.5, 1, 1.5, and 2 sigma. The red lines and the red dot mark the values of the half-mass radius and of the total mass of the cluster at that time. Histograms representing the marginalised posterior probability distribution of each parameter are also shown.}
\label{Fig_contours}
\end{figure*}

When binning data points to calculate a radial profile to be compared to theoretical predictions, information is lost that could be used to further constrain the models \citep[e.g.][]{Watkins2013}. Here we apply a discrete fitting technique to distribution function based models, using the fact that a distribution function is indeed a probability function. The comparison we propose is the first step towards applying this to observational data of globular clusters: in the case presented here the procedure is simple, due to the fact that the data we have are complete (i.e., we know all the coordinates of each star in phase space) and without observational errors.

With this approach, we do not have to build ``observational'' profiles by binning data in order to compare them with the widely used projected density and velocity dispersion profiles that can be calculated from the models. Instead we take advantage of the fact that we know the 6 coordinates of every star in phase space exactly, and we can therefore easily calculate the probability of finding a star in that position by calculating the value assumed by the distribution function of equation~(\ref{Eq_DF_Limepy}) for a given model. To do this, we compute the energy and the angular momentum of each star from its phase space coordinates; to calculate the energy, we interpolate the model potential at the position of the star.

In our fitting procedure, we use a Bayesian approach to determine the posterior probability distribution of the model parameters $\Theta$, given the data $\xvec$. The posterior probability density is given by:
\begin{equation}
p(\Theta|\xvec) = \frac{p(\Theta) p(\xvec|\Theta)}{p(\xvec)} \ ,
\end{equation}
where $p(\Theta)$ is the prior distribution, $p(\xvec|\Theta)$ is the likelihood function, and $p(\xvec)$ is the evidence, and is basically a normalisation. In this case, the data is a set of $N_{*}$ points in phase space, given by the spatial and velocity coordinates of all the stars in a snapshot. 

The likelihood function is expressed as
\begin{equation}
p(\xvec|\Theta) = \prod_{i=1}^{N_{*}} \Lambda_i(\xvec_i|\Theta) \ ,
\label{Eq_Lambda}
\end{equation}
which is the product of individual likelihood functions calculated for every star in the sample. A first obvious definition for the likelihood of individual stars is therefore exactly the distribution function:
\begin{equation}
\Lambda_i = \frac{f(\xvec_i|\Theta)}{\Lambda_0} \ ,
\end{equation}
where the parameters $\Theta$ are the 5 quantities that are necessary to identify a specific model in the \textsc{limepy} family (namely, $W_0$, $g$, $\ra$, $M$, and $\rh$, which were introduced in Section~\ref{Sect_Models}), and $\xvec_i$ is the set of phase space coordinates for the $i$-th star. The normalisation constant $\Lambda_0$ has been included such that $\int \dr^3 \rvec \, \dr^3 \vvec \, \Lambda_i = 1$. We recall that our definition of the model is such that $\int \dr^3 \rvec \, \dr^3 \vvec \, f = M$, therefore in this case $\Lambda_0 = M$.

\begin{table*}
\begin{center}
\caption[Best-fit parameters.]{Best-fit parameters. For each snapshot, identified by a label in the first column, we list the values of the best-fit parameters and of the errors, as identified by the median and 16\% and 84\% percentiles of the marginalised posterior probability distribution. Columns from 2 to 7, respectively, are as follows: the concentration parameter $W_0$, the truncation parameter $g$, the mass of the cluster $M$, the half-mass radius $\rh$, the ratio of the anisotropy radius to the half-mass radius $\ra/\rh$, the logarithm of the total mass of stars in the background $\log M_{\rm bg}$. All the quantities are expressed in \henon\ $N$-body units; the horizontal line separates snapshots taken before and after core collapse.} 
\label{Tab_BF_Param}
\begin{tabular}{ccccccc}
\hline\hline
Snapshot  & $W_0$ & $g$ & $M$ & $\rh$ & $\ra/\rh$ & $\log M_{\rm bg}$ \\
\hline
 pre-CC 1 & 4.06 $\pm$ $^{0.05}_{0.06}$ & 2.57 $\pm$ 0.02 & 0.997 $\pm$ 0.003 & 0.771 $\pm$ 0.002 & 3.14 $\pm$ 0.11 & -4.2 $\pm$ 0.2 \\
 pre-CC 2 & 4.99 $\pm$ 0.03 & 2.33 $\pm$ 0.01 & 0.998 $\pm$ 0.003 & 0.750 $\pm$ 0.002 & 2.50 $\pm$ 0.04 & -6.8 $\pm$ $^{1.3}_{1.1}$ \\
 pre-CC 3 & 5.65 $\pm$ 0.03 & 2.10 $\pm$ $^{0.02}_{0.01}$ & 0.995 $\pm$ 0.003 & 0.746 $\pm$ 0.003 & 2.31 $\pm$ 0.03 & -5.3 $\pm$ $^{0.7}_{1.7}$ \\
 pre-CC 4 & 6.22 $\pm$ 0.02 & 1.89 $\pm$ 0.01 & 0.989 $\pm$ 0.003 & 0.737 $\pm$ 0.003 & 2.20 $\pm$ 0.03 & -6.8 $\pm$ $^{1.3}_{1.2}$ \\
 pre-CC 5 & 6.61 $\pm$ 0.02 & 1.74 $\pm$ 0.01 & 0.987 $\pm$ $^{0.003}_{0.002}$ & 0.742 $\pm$ 0.003 & 2.09 $\pm$ $^{0.02}_{0.03}$ & -6.8 $\pm$ $^{1.3}_{1.2}$ \\
 pre-CC 6 & 6.98 $\pm$ 0.02 & 1.62 $\pm$ 0.01 & 0.987 $\pm$ 0.002 & 0.743 $\pm$ 0.003 & 2.00 $\pm$ 0.02 & -6.8 $\pm$ $^{1.3}_{1.1}$ \\
 pre-CC 7 & 7.31 $\pm$ 0.02 & 1.54 $\pm$ 0.01 & 0.977 $\pm$ 0.003 & 0.745 $\pm$ 0.003 & 1.93 $\pm$ 0.02 & -6.7 $\pm$ $^{1.3}_{1.2}$ \\
 pre-CC 8 & 7.61 $\pm$ 0.02 & 1.47 $\pm$ 0.01 & 0.983 $\pm$ 0.001 & 0.768 $\pm$ 0.004 & 1.83 $\pm$ 0.02 & -6.7 $\pm$ 1.3 \\
 pre-CC 9 & 8.02 $\pm$ 0.02 & 1.42 $\pm$ 0.01 & 0.968 $\pm$ $^{0.009}_{0.003}$ & 0.793 $\pm$ 0.005 & 1.69 $\pm$ 0.02 & -4.9 $\pm$ $^{0.5}_{1.0}$ \\
pre-CC 10 & 8.44 $\pm$ 0.02 & 1.40 $\pm$ 0.01 & 0.971 $\pm$ 0.002 & 0.841 $\pm$ 0.005 & 1.52 $\pm$ $^{0.02}_{0.03}$ & -6.7 $\pm$ 1.3 \\
\hline
1 & 12.22 $\pm$ 0.07 & 1.24 $\pm$ 0.01 & 0.722 $\pm$ 0.003 & 2.778 $\pm$ 0.010 &  1.01 $\pm$ 0.04 & -6.6 $\pm$ 1.2  \\  
 2 & 10.74 $\pm$ 0.04 & 1.08 $\pm$ 0.01 & 0.570 $\pm$ 0.002 & 3.972 $\pm$ $^{0.022}_{0.021}$ &  1.15 $\pm$ 0.03 & -6.8 $\pm$ 1.2  \\  
 3 & 11.37 $\pm$ 0.05 & 1.07 $\pm$ 0.01 & 0.461 $\pm$ 0.002 & 4.959 $\pm$ 0.025 &  1.38 $\pm$ 0.04 & -6.7 $\pm$ 1.2  \\  
 4 & 10.28 $\pm$ 0.05 & 0.93 $\pm$ $^{0.02}_{0.04}$ & 0.382 $\pm$ 0.002 & 5.584 $\pm$ $^{0.040}_{0.038}$ &  1.85 $\pm$ $^{0.07}_{0.08}$ & -4.3 $\pm$ 0.3  \\  
 5 & 10.19 $\pm$ 0.05 & 0.93 $\pm$ 0.01 & 0.313 $\pm$ 0.001 & 5.922 $\pm$ 0.037 &  2.89 $\pm$ $^{0.22}_{0.18}$ & -4.7 $\pm$ $^{0.3}_{0.4}$  \\  
 6 & 11.75 $\pm$ 0.08 & 0.97 $\pm$ 0.02 & 0.249 $\pm$ 0.001 & 6.093 $\pm$ $^{0.039}_{0.040}$ &  3.57 $\pm$ $^{0.53}_{0.37}$ & -4.7 $\pm$ $^{0.3}_{0.4}$  \\  
 7 &  9.88 $\pm$ 0.05 & 0.75 $\pm$ 0.01 & 0.187 $\pm$ 0.001 & 5.980 $\pm$ $^{0.038}_{0.040}$ & 40.8 $\pm$ $^{\infty}_{26.6}$ & -4.3 $\pm$ 0.2  \\  
 8 &  9.42 $\pm$ 0.06 & 0.77 $\pm$ 0.01 & 0.136 $\pm$ 0.001 & 5.182 $\pm$ $^{0.056}_{0.055}$ & 72.1 $\pm$ $^{\infty}_{44.0}$ & -6.5 $\pm$ 1.3  \\  
 9 & 10.53 $\pm$ 0.10 & 0.86 $\pm$ 0.02 & 0.088 $\pm$ 0.001 & 4.871 $\pm$ $^{0.053}_{0.052}$ & 38.8 $\pm$ $^{\infty}_{23.1}$ & -6.7 $\pm$ 1.2  \\  
10 & 10.35 $\pm$ 0.13 & 0.81 $\pm$ 0.03 & 0.048 $\pm$ 0.001 & 4.017 $\pm$ $^{0.061}_{0.062}$ & 38.4 $\pm$ $^{\infty}_{23.4}$ & -6.8 $\pm$ 1.2  \\  
11 &  8.60 $\pm$ 0.18 & 0.48 $\pm$ 0.05 & 0.016 $\pm$ 0.001 & 2.798 $\pm$ $^{0.096}_{0.100}$ & 63.7 $\pm$ $^{\infty}_{48.1}$ & -6.7 $\pm$ 1.2  \\
\hline
\end{tabular}
\end{center}
\end{table*}

This formulation, however, has some problems. As stated in equation~(\ref{Eq_DF_Limepy}), the distribution function vanishes when $E\geq\phi(\rt)$, therefore whenever we have a single star that does not fulfil this requirement $p(\xvec|\Theta) = 0$, from equation~(\ref{Eq_Lambda}). This hard cut-off is perhaps not realistic, because the models may not be perfect: to take this into account, therefore, we decided to allow for the possibility of having a non-zero number of stars in the system that cannot be described by the best-fit \textsc{limepy} model. To do this, we decided to add a constant background to the model, so that the likelihood is always non-zero, even for combinations of model parameters that correspond to the case described above:
\begin{equation}
\Lambda_i = \left[f(\xvec_i|W_0,g,\ra,M,\rh) + \frac{M_{\rm bg}}{V_{*}} \right] \frac{1}{\Lambda_0} \ .
\label{Eq_Lambda_i}
\end{equation}
The second term in the equation represents the mass density of a uniform background of stars with total mass $M_{\rm bg}$ that extends on the entire volume in the phase space that is occupied by the stars, $V_{*}$. The normalisation constant is now given by
\begin{equation}
\Lambda_0 = M + M_{\rm bg} \ .
\end{equation}
We consider $M_{\rm bg}$ to be another fitting parameter, which represents the number of stars that are not described by the \textsc{limepy} models defined in equation~(\ref{Eq_DF_Limepy}): the smaller it is, the better the model reproduces the data. The total number of fitting parameters is therefore 6, and $\Theta = \lbrace W_0,g,\ra,M,\rh, M_{\rm bg} \rbrace$. We point out that although we know the total number of stars in each snapshot, we are not using this information in the computation of the mass $M$. This is the reason why in principle it would possible to determine all the best-fit parameters by considering only a subset of the stars, and why it could be possible to obtain a best-fit value of the mass larger than the true one calculated from the snapshot.

For the parameters, we choose to use uniform priors over the following ranges: $4\!<\!W_0\!<\!15$, $0.3\!<\!g\!<\!2.1$, $0.001\!<\!M\!<\!1.5$, $0.2\!<\!r_{\rm h}\!<\!15$, $-1\!<\!\log \ra\!<\!3.7$, and $-8.5\!<\!\log M_{\rm bg}\!<\!-1$. We consider $\log \ra$ as a fitting parameter instead of $\ra$ in order to have an uninformative prior for this parameter since it can span several orders of magnitude. We also consider a logarithmic value to fit on $M_{\rm bg}$, because it usually assumes very small values.

We use a Markov chain Monte Carlo fitting technique to explore the parameter space and to efficiently sample the posterior probability distribution for the parameters above. We use \textsc{emcee} \citep{emcee_paper}, a \textsc{python} implementation of Goodman \& Weare's affine invariant Markov chain Monte Carlo ensemble sampler\footnote{\textsc{emcee} is available online at \url{https://github.com/dfm/emcee}}. Typically, we consider 200 walkers, each of which takes 1000 steps in parameter space. The fact that \textsc{limepy} models are very fast to solve allows us to carry out this fitting procedure for each snapshot in $\sim 2$ hours. We initialise the walkers by putting them all in a sphere in parameter space, with a spread of $\Theta_i \times 10^{-3}$ around the starting value we chose for each parameter $\Theta_i$.

In Fig.~\ref{Fig_contours} we show the 2d-projections of the posterior probability distribution on the planes determined by every pair of parameters that we obtained as a result of the fit carried out with \textsc{emcee} for the snapshot number 5 after core  collapse, taken at $t = 2.5 \times 10^5$ $N$-body times. From Fig.~\ref{Fig_contours} we can see that there is a slight degeneracy between the truncation parameter and the anisotropy radius, that is also observed for the other snapshots considered. Indeed, when considering the extent of the model, increasing $g$ or decreasing $\ra$ have the same effect as increasing the truncation radius. Also shown in Fig.~\ref{Fig_contours} are the histograms representing the marginalised posterior probability distribution for each parameter. 

Typically, convergence is obtained after a burn-in phase of about 150 steps. We ran the same fit multiple times to ensure that by changing the initial position of the walkers the result did not change. This is necessary because in some cases it could happen that walkers starting in a certain position of the parameter space get momentarily trapped in a local maximum of the likelihood, and they eventually converge only after a very long time (more than 1000 steps). We also tried to spread the walkers in the parameter space in different ways, which also had no detectable change in the final result.

Finally, we wish to perform a detailed comparison between the \textsc{limepy} models and the more commonly adopted isotropic \cite{King1966} models. To do this, we carry out an additional fit to all the snapshots, by imposing the value of $g = 1$, corresponding to the truncation prescription of King models, and by considering $\ra = \infty$, to have isotropic models. In this case, we are therefore left with only 4 fitting parameters: $\Theta = \lbrace W_0,M,\rh, M_{\rm bg} \rbrace$. We present the results relative to this approach in Section~\ref{Sect_Results_3}.

\begin{figure*}
\includegraphics[width=\textwidth]{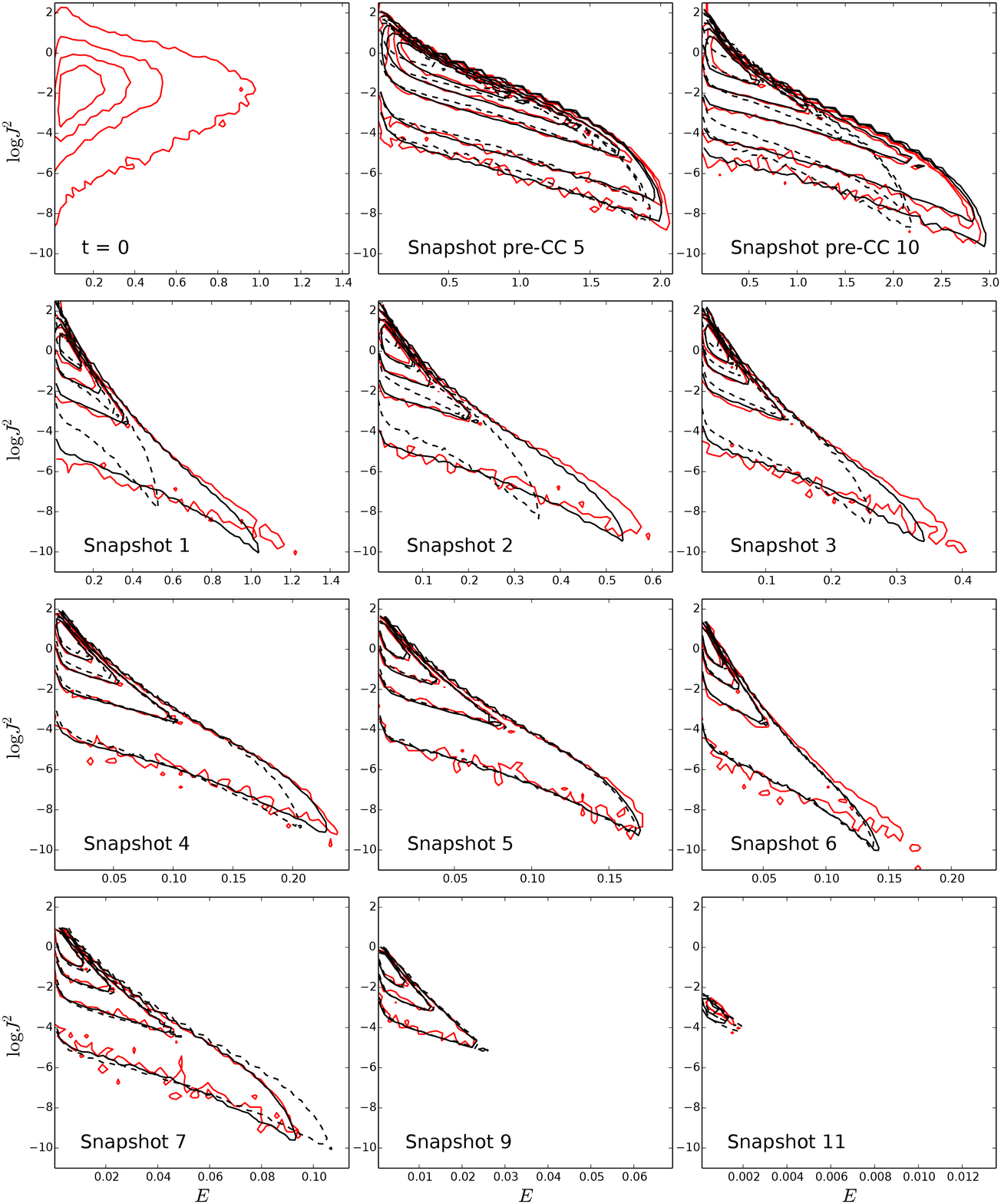}
\caption{Contours of the density of stars $N(E,J^2)$ in the plane identified by the energy $E$ and angular momentum $J$, expressed in \henon\ $N$-body units. Each panel corresponds to a given snapshot, with labels listed in Table~\ref{Tab_True_Properties}. The red lines refer to the distribution of the stars, as calculated for stars in the $N$-body snapshots; solid and dashed black lines are the contours of the best-fit \textsc{limepy} and King model,respectively. The contours shown enclose 25\%, 50\%, 75\%, and 97\% of the total mass of the cluster at each moment; in the case of Snapshot 9 we omit the contour corresponding to 97\% of the mass, and for Snapshot 11 we show the contours enclosing 10\% and 15\% of the total mass. Each panel in the figure has a different range in energy, to better show the agreement between the models and the snapshots.}
\label{Fig_densityEJ}
\end{figure*}

\section{Results and discussion}
\label{Sect_Results}

We describe here the main results of our analysis. In Sect.~\ref{Sect_Results_0} we propose a phase space comparison of best-fit models and data from the simulation, in Sect.~\ref{Sect_Results_1} we compare the radial profiles of moments of the distribution function defining the family of \textsc{limepy} models to the data from the snapshots under consideration, in Sect.~\ref{Sect_Results_2} we discuss the evolution of the values of best-fit parameters in time, and in Sect.~\ref{Sect_Results_3} we compare the results obtained with \textsc{limepy} models to those obtained with King models.

Table~\ref{Tab_BF_Param} lists the values of the best-fit parameters and the respective errors. The best-fit value is identified by taking the median value of the correspondent marginalised posterior probability distribution, and the errors correspond to the 16\% and 84\% percentiles; different values for the upper and lower error are obtained when the distribution is not symmetric. We decide to consider here the ratio of the anisotropy radius to the half-mass radius, $\ra/\rh$, instead of the absolute value of $\ra$, to better represent the amount of anisotropy relative to the specific structure of the system. We compute this ratio for all the steps and for all the walkers, and then we compute the median and the 16\% and 84\% percentiles of this marginalised distribution. Note that we do not fit on the ratio $\ra/\rh$ because the value of $\rh$ for a model is obtained only after the model has been calculated and therefore using this as a fitting parameter would require several iterations, and would be computationally expensive.

\begin{figure*}
\includegraphics[width=\textwidth]{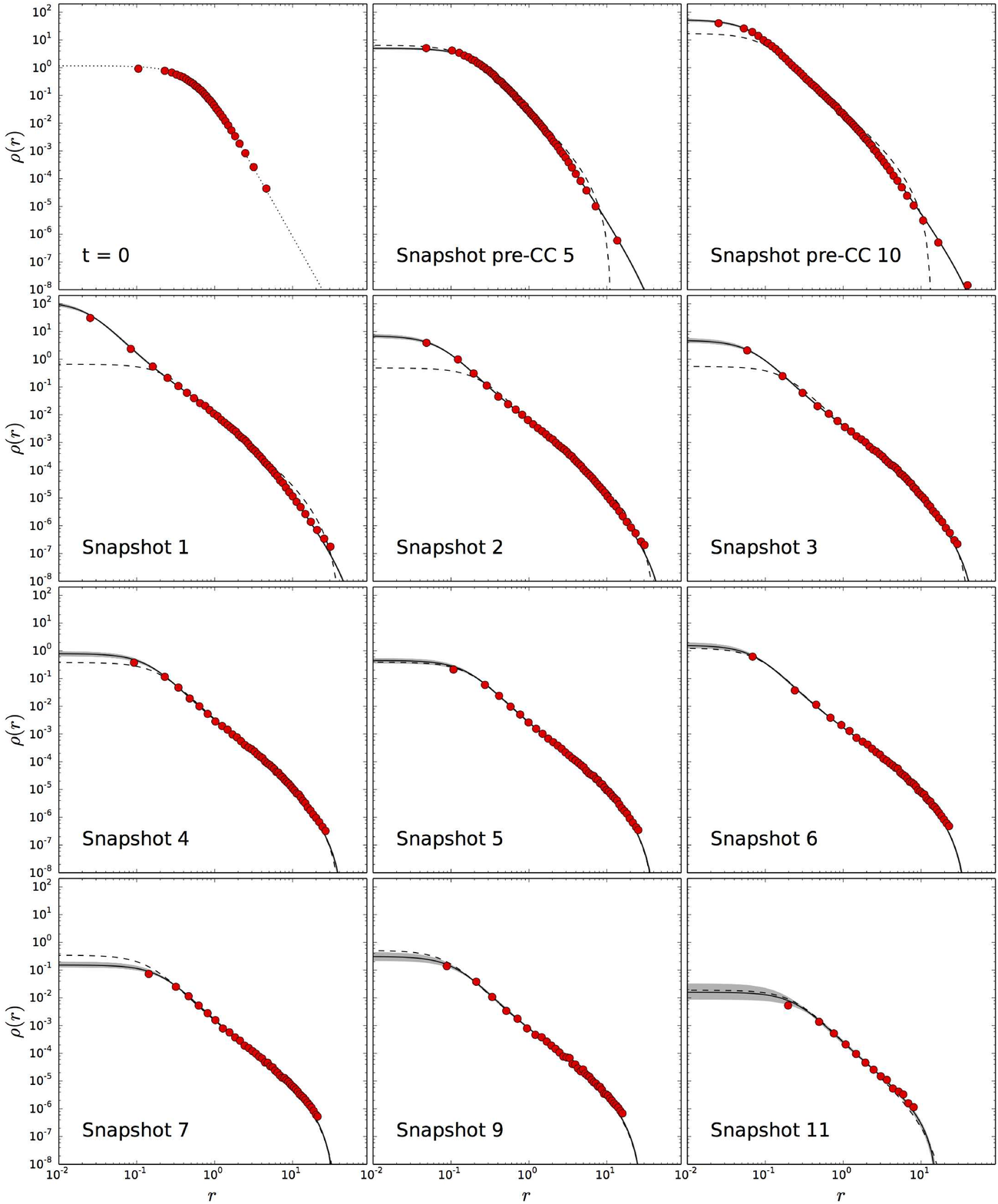}
\caption{Mass density profile of the cluster at different times. Each panel corresponds to a given snapshot, with labels listed in Table~\ref{Tab_True_Properties}. Density and radial coordinate are expressed in \henon\ $N$-body units. Solid lines represent the density profiles $\rho$ calculated from the best-fit \textsc{limepy} models, the red dots those calculated from the data; error bars are also shown. Grey lines indicate the profiles calculated for 200 models randomly selected among those explored by the \textsc{emcee} chains. Dashed lines indicate the best-fit King model profiles. The top left panel corresponds to the sample of stars used as initial condition to start the simulation, and generated from a Plummer model: the dotted line in this panel represents the Plummer model theoretical density profile, and the red dots the profile calculated by binning the stars.}
\label{Fig_dens}
\end{figure*}

\begin{figure*}
\includegraphics[width=\textwidth]{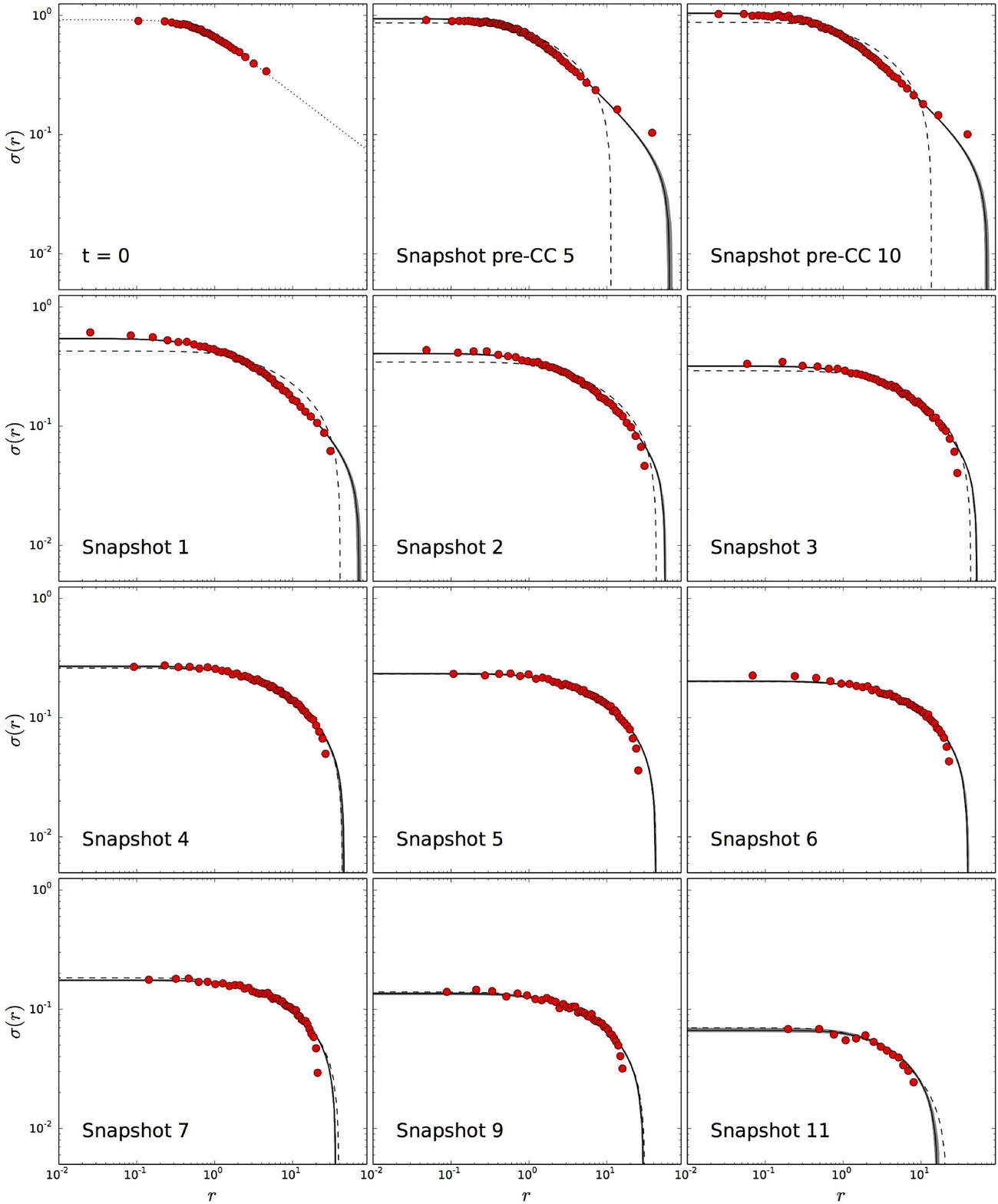}
\caption{Velocity dispersion profile of the cluster at different times. Each panel corresponds to a given snapshot, with labels listed in Table~\ref{Tab_True_Properties}. Velocity dispersion and radial coordinate are expressed in \henon\ $N$-body units. Solid lines represent the velocity dispersion profiles $\sigma$ calculated from the best-fit \textsc{limepy} models, the red dots those calculated from the data; error bars are also shown. Grey lines indicate the profiles calculated for 200 models randomly selected among those explored by the \textsc{emcee} chains. Dashed lines indicate the best-fit King model profiles. The top left panel corresponds to the sample of stars used as initial condition to start the simulation, and generated from a Plummer model: the dotted line in this panel represents the Plummer model theoretical velocity dispersion profile, and the red dots the profile calculated by binning the stars.}
\label{Fig_disp}
\end{figure*}

\begin{figure*}
\includegraphics[width=\textwidth]{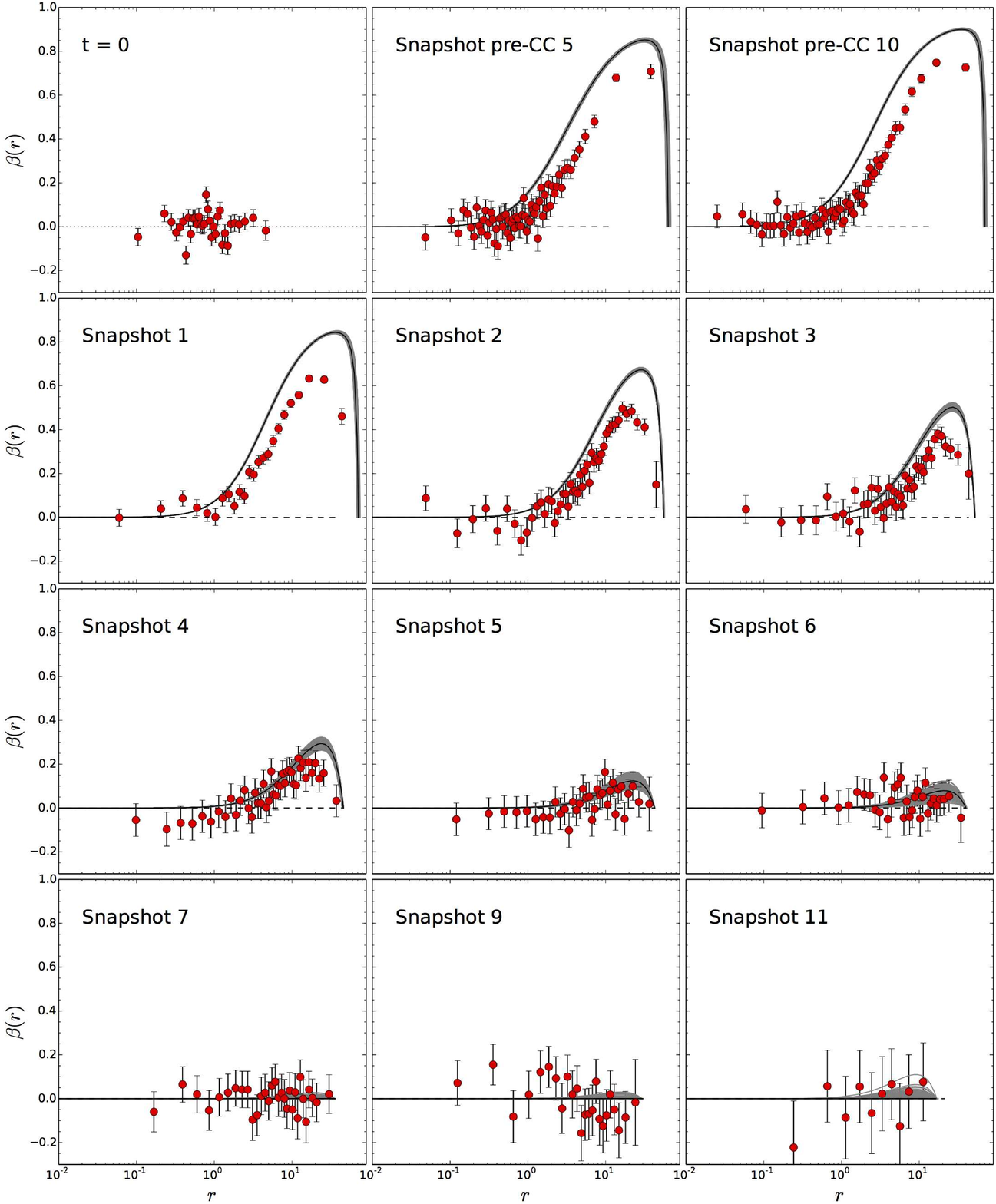}
\caption{Anisotropy profile of the cluster at different times. Each panel corresponds to a given snapshot, with labels listed in Table~\ref{Tab_True_Properties}. The radial coordinate is expressed in \henon\ $N$-body units. Solid lines represent the anisotropy profiles $\beta$ calculated from the best-fit  \textsc{limepy} models, the red dots those calculated from the data; error bars are also shown. Grey lines indicate the profiles calculated for 200 models randomly selected among those explored by the \textsc{emcee} chains. Dashed lines indicate the best-fit King model profiles. The top left panel corresponds to the sample of stars used as initial condition to start the simulation, and generated from a Plummer model: the dotted line in this panel represents the initial anisotropy profile, showing that the initial condition is isotropic, and the red dots the profile calculated by binning the stars.}
\label{Fig_anis}
\end{figure*}

\subsection{Phase space assessment}
\label{Sect_Results_0}

As we pointed out in the previous section, we applied a discrete fitting method to determine the best-fit parameters of the models. On the one side, this is very useful, because it allows us to retain all the information the data can provide. On the other side, however, it is not easy to visualise the goodness of a fit. Therefore, to explore the agreement between the best-fit models and the numerical simulation under consideration, we have performed a detailed comparison in phase space, which is accessible, once again, by virtue of the full dynamical information provided by the synthetic data at our disposal. As appropriate for the description of any spherical (anisotropic) stellar system, we refer to the partition $(E,J^2)$ and we consider the density $N(E,J^2)$, such that $M=\int {\rm d} E \, {\rm d}J^2 N(E,J^2)$, where $M$ is the total mass of the cluster. We have calculated such a density both from selected snapshots of our reference $N$-body model, and from the corresponding best-fit \textsc{limepy} and King models \citep[by means of a sampling of the distribution function with Monte Carlo techniques, with $10^6$ particles; for details, see][]{GielesZocchi2015}.

Figure~\ref{Fig_densityEJ} shows the contours of the density of stars $N(E,J^2)$ in the plane identified by energy and angular momentum. Each panel in the figure corresponds to a given snapshot (as indicated in the plots; see also Table~\ref{Tab_True_Properties}). We decided to represent here only some of the available snapshots, to highlight the most significant variations associated with the main stages of evolution of the cluster. Red lines refer to the actual distribution of the stars of each snapshot, solid black lines are the contours of the best-fit \textsc{limepy} model, and dashed black lines those of the best-fit King model for each snapshot. The contours shown enclose 25\%, 50\%, 75\%, and 97\% of the total mass of the cluster at each moment. In the case of Snapshot 9 we omit the contour corresponding to 97\% of the mass, and for Snapshot 11 we show the contours enclosing 10\% and 15\% of the total mass, because at this stage the cluster contains too few stars, and the contours enclosing larger fraction of the mass have highly irregular shape. Each panel in the figure has a different range in energy, to better show the agreement between the models and the snapshots.

Unlike the radial profiles we discuss in Sect.~\ref{Sect_Results_1}, here we take into account the stars discretely, as we did in the fitting procedure. Moreover, the density $N(E,J^2)$ provides a more direct comparison between the quantities that are used in the fitting procedure: indeed, energy and angular momentum represent the way the phase space coordinates are accounted for in the distribution function. For the models, the energy is calculated by considering the potential of the model as a function of the distance to the centre of the cluster. When considering the $N$-body snapshots, we calculate the energy of each star by using the value of the true potential energy of the stars in the cluster, as calculated from the snapshots. When fitting the models to the snapshot we do not consider the true potential of the cluster, because we only want to consider the 6-dimensional coordinates of the phase space for each star. In this respect, Fig.~\ref{Fig_densityEJ} is an opportunity to test how well the model potential describes the actual cluster potential.

We wish to emphasise that the overall agreement is particularly good, not only in the proximity of the maximum of the distribution of the synthetic data in phase space, but also within the lower density regions, in which the behaviour of the simulation particles is captured very well by the \textsc{limepy} best-fit distribution function. The comparison with King models is particularly instructive, as it results that, especially in the proximity of the core collapse, they offer a very approximate description of the $N$-body simulation (see Sect.~\ref{Sect_Results_3} for further discussion).

\subsection{Comparison of radial profiles}
\label{Sect_Results_1}

Driven by similar motivations, we have also calculated the three-dimensional density, velocity dispersion, and anisotropy profiles for the simulation snapshots by binning the data in spherical shells containing an equal number of particles. The number of stars in each bin (and the total number of bins) varies for different snapshots, and it has been chosen in order to have profiles that are both rich and accurate (the number of stars per bin varies from $\approx$ 1000 for the early snapshots to $\approx$ 100 for the late ones).

Figures~\ref{Fig_dens}, \ref{Fig_disp}, and \ref{Fig_anis} show the density, velocity dispersion, and anisotropy profiles, respectively, for the same snapshots illustrated in Fig.~\ref{Fig_densityEJ}. The solid lines represent the profiles calculated from the best-fit \textsc{limepy} models, the red dots are the ones calculated from the data. Error bars are also shown, but in some cases they are not visible, because they have a size smaller than that of the dots. We also show, as grey lines, the profiles calculated for 200 models randomly selected among those explored by the \textsc{emcee} walkers in the post burn-in phase: this is to give an idea of the uncertainty in the best-fit models and corresponding profiles. The scales of the plots are the same for all the snapshots, to make it easier to compare how the profiles are changing with the evolution of the cluster. In each panel, a dashed line reproduces the best-fit King model profile for each quantity: we postpone a discussion on these models, and on the comparison to the results obtained with the \textsc{limepy} models, to Section~\ref{Sect_Results_3}.

The top left panel in each of these figures corresponds to the sample of stars generated from the spherical and isotropic \cite{Plummer1911} model used as initial condition to start the simulation. The dotted lines represent the quantities calculated directly from the equations defining the model. We also overplot the profiles calculated by binning the data generated from the model, in a similar way as we did for the other snapshots. 

\subsubsection{Density and velocity dispersion profiles}

When looking at different snapshots, we notice immediately that before core collapse the density increases in the centre, and then it decreases, as a function of time, 
because of expansion and mass loss driven by two-body relaxation (see Fig. ~\ref{Fig_dens}).
As illustrated by the grey lines in the figure, the largest uncertainties on the best-fit profiles are found in the innermost parts, for the latest snapshots. This is due to the fact that, with time, the cluster becomes less concentrated, and less stars are found in its centre: the models are therefore less constrained in that radial range, while for the rest of the profile all the models overlap, and the differences from one another are very small. We point out that the best-fit \textsc{limepy} models are able to reproduce the density profiles remarkably well, out to their outermost parts, for all the considered snapshots.

It is immediately clear, from a comparison of the different panels, that after an initial expansion in the pre-collapse phase, the truncation radius becomes smaller as time passes. The truncation appears to be more shallow (large $g$) at the beginning, and it becomes steeper (small $g$) during the evolution, as particularly evident when inspecting the density profiles in Fig.~\ref{Fig_dens}.

By inspecting Fig.~\ref{Fig_disp}, we notice that also the velocity dispersion first increases, in the pre-collapse phase, and then decreases in time. The \textsc{limepy} models seem to be adequate in reproducing the profiles calculated from the data, even though there are some discrepancies at large radii. Indeed, we note that the velocity dispersion profiles of the best-fit models at early times underestimate in the outer parts the ones calculated from the snapshots, while at later times they overestimate them. This is particularly evident in the very final stages of evolution, from snapshots 7 onwards. The shape of the profiles also changes, becoming overall more shallow. 

\begin{figure*}
\includegraphics[width=0.49\textwidth]{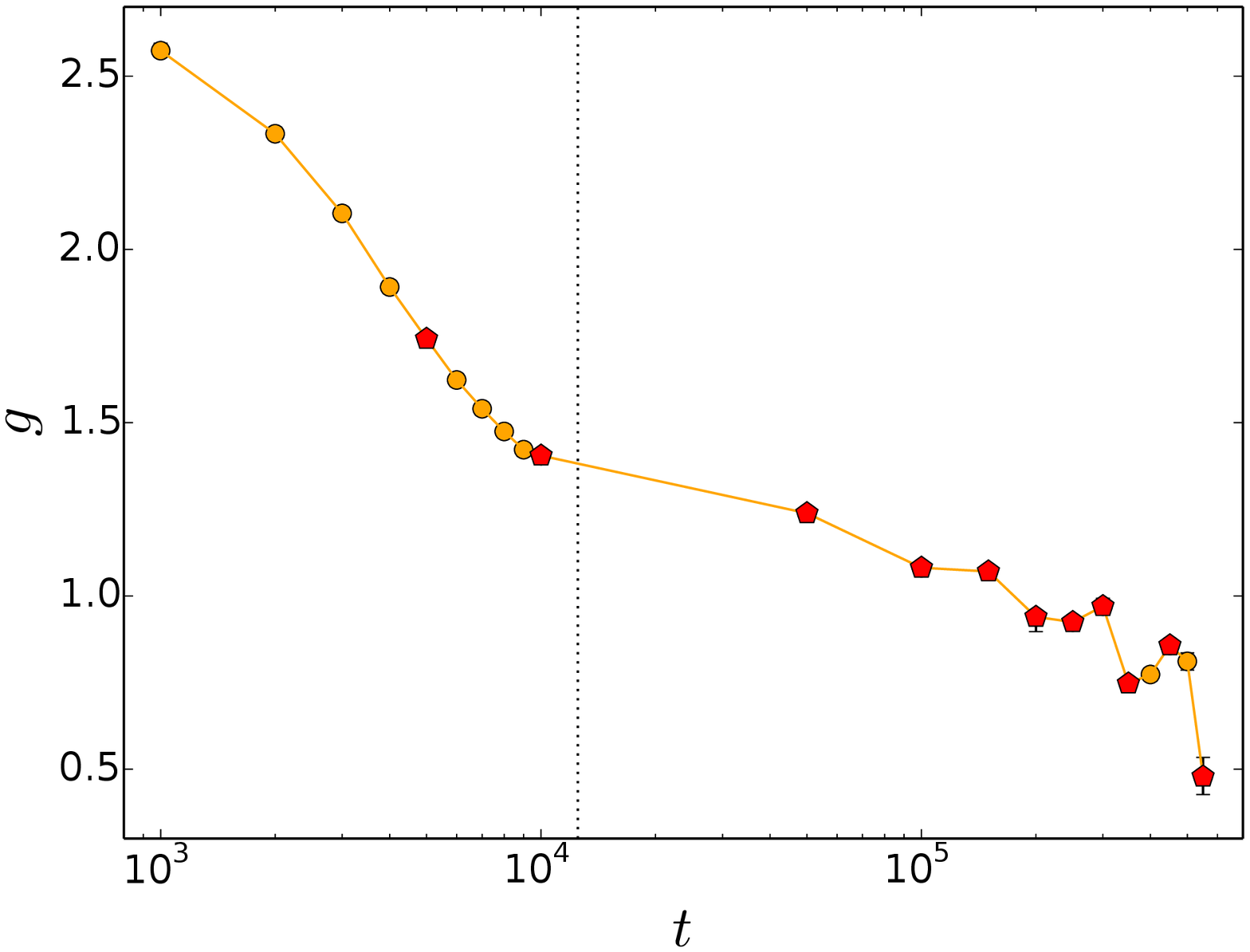} \quad
\includegraphics[width=0.49\textwidth]{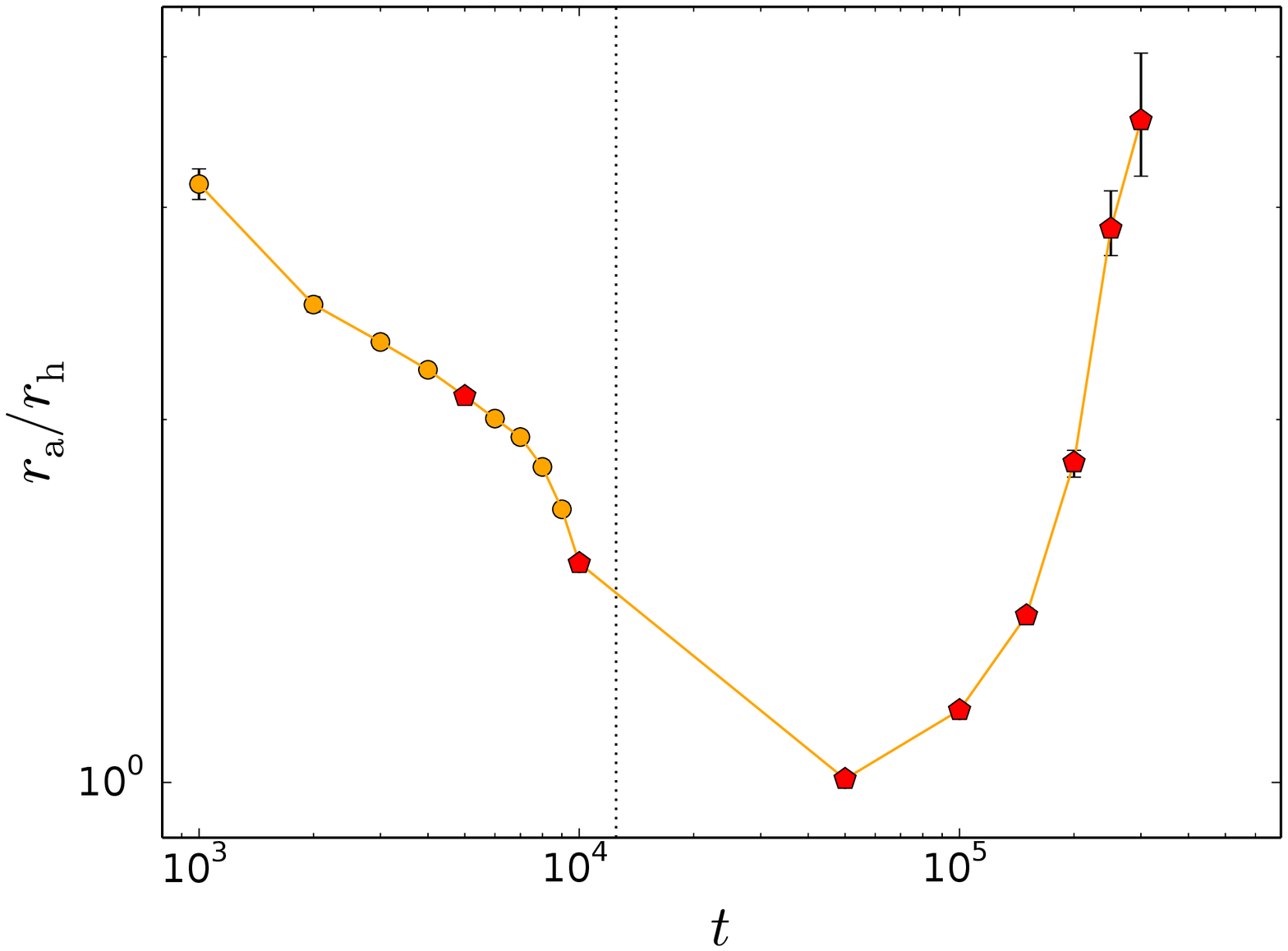}
\caption{Time evolution of best-fit truncation and anisotropy parameters. On the left we show the evolution of $g$, on the right of the ratio of anisotropy radius to half-mass radius $\ra/\rh$. Orange dots with error bars represent the best-fit values obtained for the parameters for each snapshot (see Table~\ref{Tab_BF_Param}); an orange line connecting the dots has been added to better show the trend. Moreover, we mark with red pentagons the values corresponding to the snapshots whose relevant profiles are reproduced in Figs.~\ref{Fig_densityEJ}--\ref{Fig_anis}. In the right panel we exclude snapshots 7 to 11 after core collapse, because the cluster at that point is essentially isotropic. The vertical dotted lines indicate the moment of core collapse.}
\label{Fig_g_ra}
\end{figure*}

\subsubsection{Anisotropy profiles}
\label{Sect_Results_1_anis}

In Fig.~\ref{Fig_anis} we show the anisotropy profiles for the cluster at different times. As stated in Section~\ref{Sect_Simulations} and shown in the first panel of the figure, the initial conditions of the cluster are isotropic. Then, the anisotropy profile changes significantly in time, and its evolution can be divided into two parts, separated by the core collapse.

In the pre-collapse phase, the cluster develops an increasingly large degree of radial anisotropy: it appears to be isotropic in the centre and radially anisotropic in the outer parts. As time passes, the width of the profile increases, its maximum rises in value, and moves outwards. The development of radial anisotropy is related to the fact that, during the early evolution, stars are scattered outside the core in radial orbits, and this process contributes to increase the radial component of the velocity dispersion.

After core collapse, the degree of anisotropy decreases. The profiles are isotropic close to the truncation radius, beyond the region characterised by radial anisotropy, the extent of which decreases in time, until the entire cluster becomes again isotropic. This happens as a result of two main effects. First, mass loss through dynamical processes, which is enhanced by the presence of the tidal field with respect to an isolated case, is removing the outer, more radially anisotropic layers of the cluster \citep{GierszHeggie1997}. Second, the tidal torque induces isotropy in the velocity dispersion of the outer regions of the cluster \citep{OhLin1992}.

We notice that the anisotropy profile calculated by binning the stars in snapshot 6 is not very different from the one calculated for snapshot 7, but at this point the selected best-fit models make the transition from the modestly anisotropic to the fully isotropic regime (this is particularly evident when inspecting the values of the ratio $\ra/\rh$ listed in Table~\ref{Tab_BF_Param}): this transition happens at approximately the same time at which the cluster becomes tidally filling (i.e., when $\rh/\rj \sim 0.13$, as stated in Section~\ref{Sect_Simulations}). 

By inspecting Fig.~\ref{Fig_anis}, it is particularly evident that, especially for the earlier snapshots, there appears to be a discrepancy between the best-fit models and the data in proximity of the peak of the anisotropy profile, with the model overestimating the degree of anisotropy (see also Appendix~\ref{App_residuals} for a more detailed discussion). The exploration of the behaviour of the models in the phase space (see Sect.~\ref{Sect_Results_0}) offers partial reassurance on this point, as it helps clarifying the origin of the presence of the anisotropy in the velocity space, in connection with the role of angular momentum. In fact, especially immediately after core collapse (see panels corresponding to Snapshots 1, 2, 3 in Fig.~\ref{Fig_densityEJ}), the most significant discrepancy in phase space between the $N$-body simulation and the best-fit \textsc{limepy} models can be identified in the region corresponding to stars with relatively high energy and low angular momentum. In such a regime, \textsc{limepy} models tend to favour even slightly lower values of $J$ (i.e., the black contours are lower than the red ones), which is directly linked to the tangential velocity component $v_{\rm t}$, which, in turn, determines a systematically higher value of radially-biased anisotropy, as illustrated in $\beta$ radial profiles in the corresponding panels of Fig.~\ref{Fig_anis}. In the pre-collapse phase, such a behaviour is still present, although to a smaller extent, while the regime in which the discrepancy between \textsc{limepy} models and the simulation seems to be more significant is at very low energies (i.e., very bound stars) with intermediate to high values of angular momentum; in this case the interpretation is less straightforward as the behaviour of the models with respect to the simulations is mixed.

We recall that the anisotropy is the most uncertain quantity among the ones considered here. Indeed, this is evident when looking at the size of the error bars for the points in the profile, as compared to those obtained for the density and velocity dispersion profiles. We point out that the characterisation of the anisotropy profile in the final stages of evolution is particularly challenging with respect to the initial snapshots, because of the relatively weak deviations from isotropy and because of the reduced number of particles in the simulation. A quantitative discussion about comparison of the radial profiles, as resulting from best-fit models and the reference $N$-body simulation, of the observables discussed in this section is presented in Appendix~\ref{App_residuals}.

\subsection{Evolution of model parameters}
\label{Sect_Results_2}

\subsubsection{Anisotropy and tides}

It is particularly interesting to inspect the evolution of the anisotropy radius $\ra$ and the truncation parameter $g$, because their behaviour gives us some insights on the role played by radially-biased anisotropy and tides in determining the internal dynamics of the cluster. Fig.~\ref{Fig_g_ra} shows the evolution of these two parameters. 

\begin{figure*}
\includegraphics[width=0.49\textwidth]{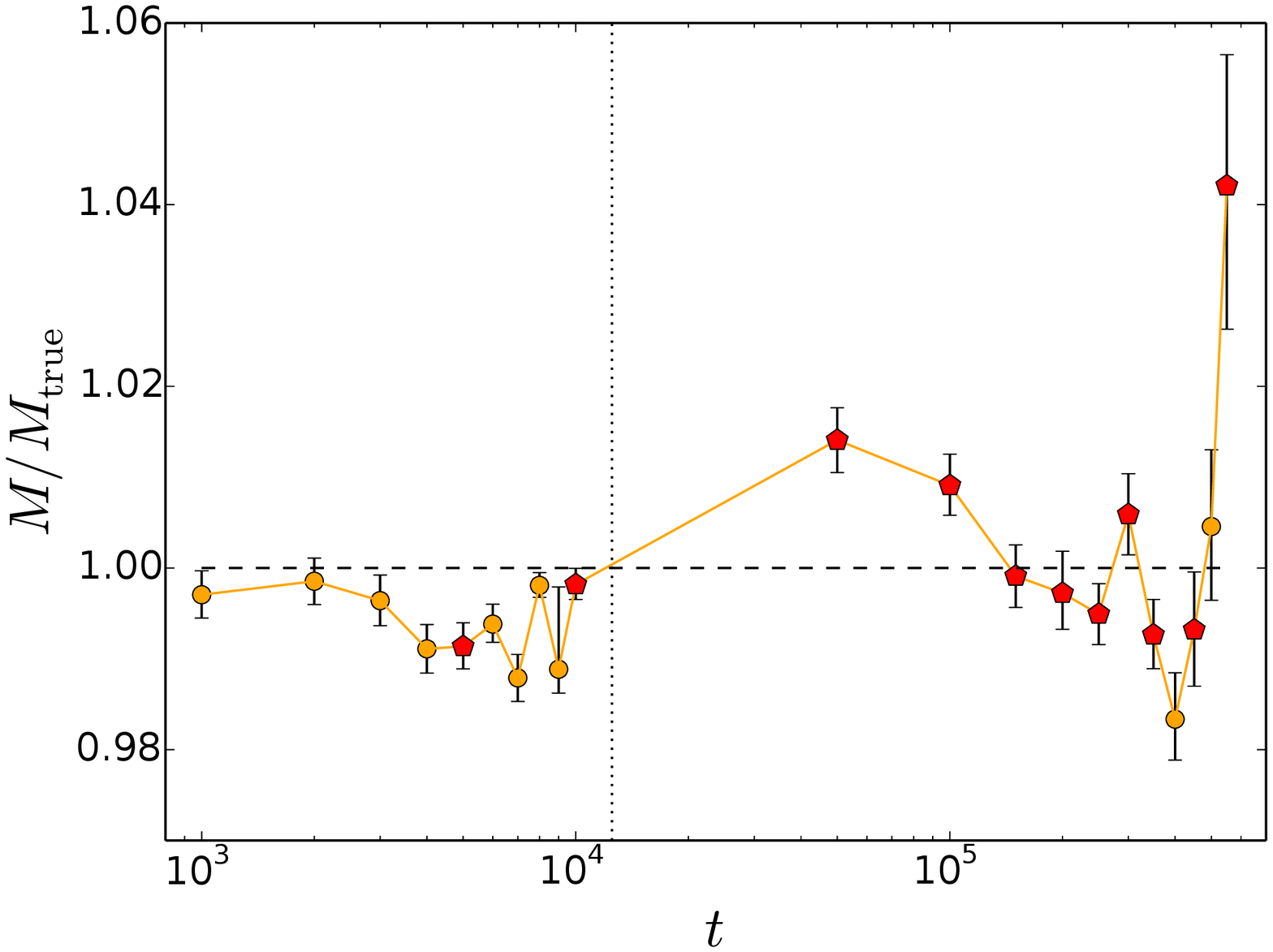} \quad
\includegraphics[width=0.49\textwidth]{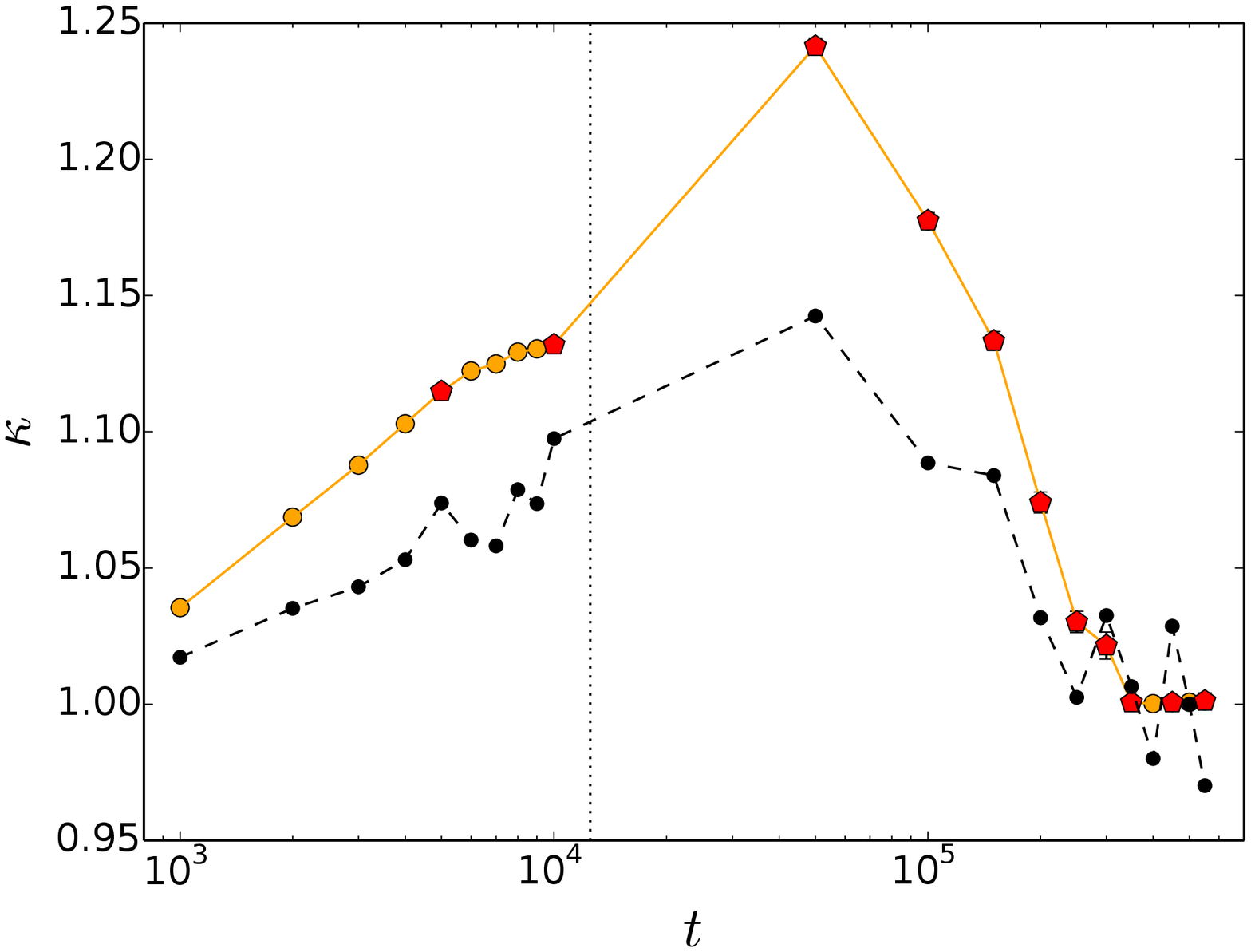} \\
\includegraphics[width=0.49\textwidth]{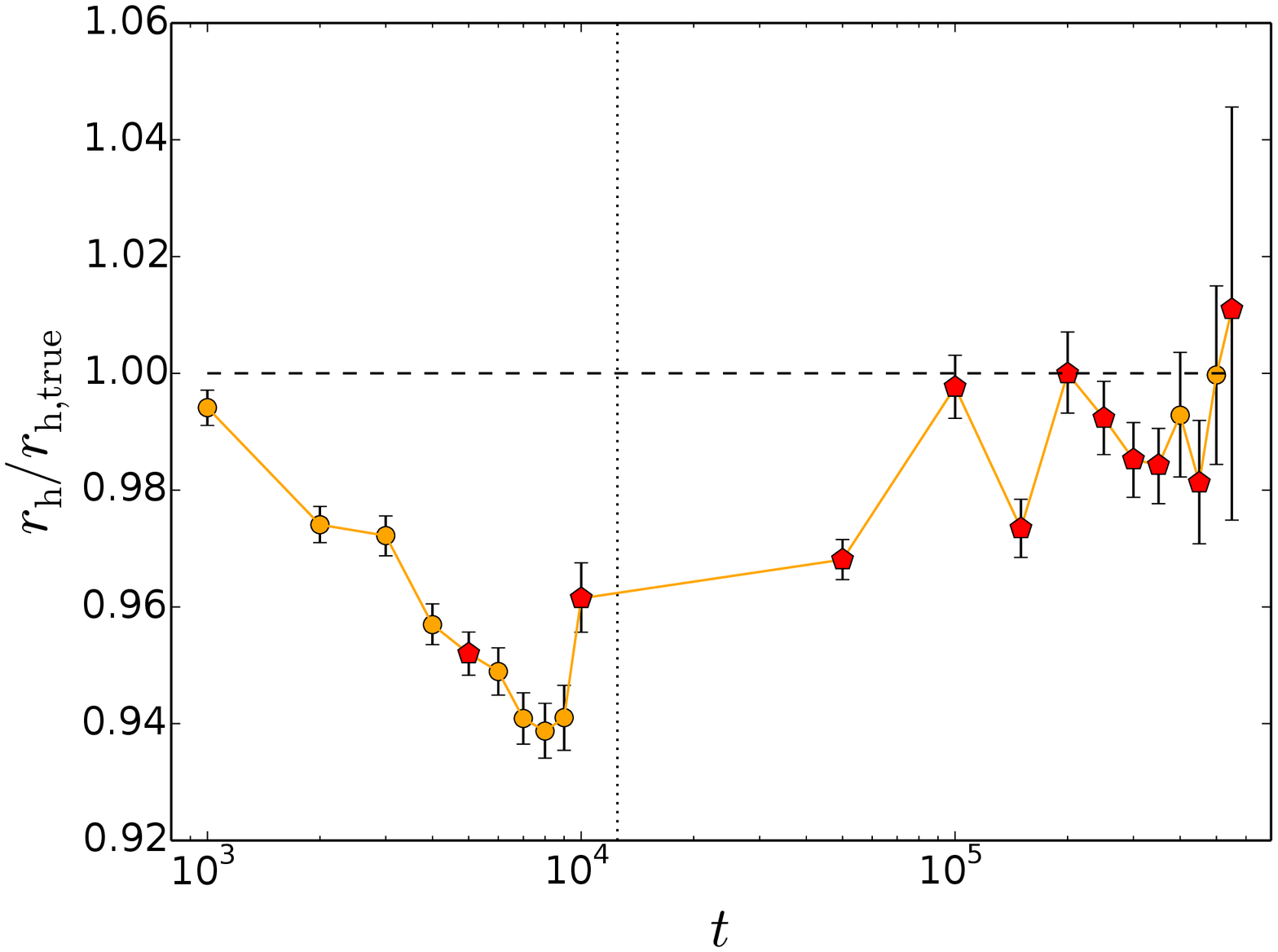} \quad
\includegraphics[width=0.49\textwidth]{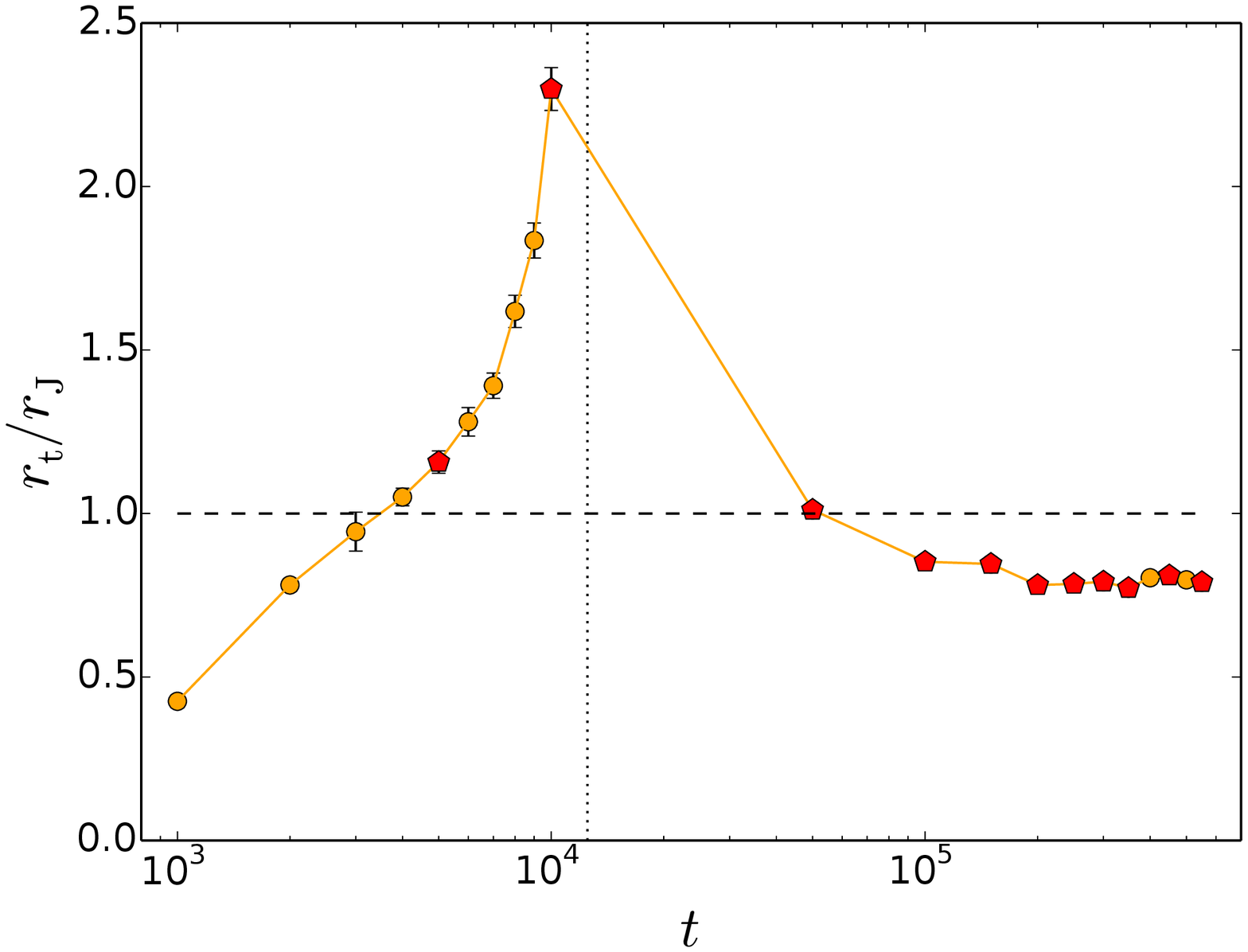}
\caption{Time evolution of best-fit values for some relevant quantities. On the left, we show the ratio of best-fit values (see Table~\ref{Tab_BF_Param}) to the true values obtained directly from the snapshots (see Table~\ref{Tab_True_Properties}) for the mass $M$ (top panel) and for the half-mass radius $\rh$ (bottom panel). The bottom right panel shows the ratio of the truncation radius $\rt$ of the best-fit \textsc{limepy} models to the true value of the Jacobi radius of the cluster. Orange dots represent the values of the ratios obtained for each snapshot; error bars are also shown. In these panels, the dashed lines mark the position of the unity. The top right panel shows the evolution of the values of the anisotropy parameter $\kappa$ as computed from the snapshot and as calculated for the best-fit \textsc{limepy} models, indicated with black and orange dots, respectively. Moreover, in each panel we mark with red pentagons the values corresponding to the snapshots whose relevant profiles are reproduced in Figs.~\ref{Fig_densityEJ}--\ref{Fig_anis}. The vertical dotted lines indicate the moment of core collapse.}
\label{Fig_properties}
\end{figure*}

As shown in the left panel in the figure, the truncation parameter $g$ decreases in time, from $\sim 2.5$ down to $\sim 0.5$. This behaviour reflects the fact that, during the evolution, the role of tides becomes more important in shaping the structure of the cluster, which is progressively filling its Roche volume, with the truncation being more abrupt at the end of the evolution. We recall here that a value of $g = 0$ corresponds to models with the same truncation prescription as \cite{Woolley1954} models, $g = 1$ to models with the same truncation as \cite{King1966} models, and $g = 2$ to models with the same truncation as \cite{Wilson1975} models. Observational studies \citep[e.g.,][]{MLvdM2005} seem to indicate that the preferred truncation prescription is Wilson-like (i.e. $g=2$). A possible interpretation of this is that a large fraction of Galactic globular clusters are still in the early phases of evolution. If clusters formed dense (i.e. high ratio of $\rj/\rh$), they spend roughly the first half of their evolution expanding towards their tidal bound \citep{GielesHeggieZhao2011}. As long as the ratio $\rt/\rh \gtrsim 10$, King models ($g=1$) are unable to describe the outer parts of the cluster \citep{Baumgardt2010}, and this may be why Wilson models ($g=2$) are preferred in the study by \cite{MLvdM2005}. We note that a direct comparison between the value of $g$ of the equal-mass models in this study and real globular clusters should be done with caution, because the $N$-body model discussed here reaches much higher central densities in core collapse than real clusters with a mass spectrum, and it does not capture the effect of mass segregation. Also, we only considered the bound stars, and this choice likely leads to smaller values of $g$ because between $0.8\rj$ and $\rj$ most stars are energetically unbound. And finally, we find that $g$ decreases when the cluster reaches core collapse, which in real globular clusters can take much longer because of the effect of primordial binaries \citep[][]{VesperiniChernoff1994,Trenti2007}, stellar evolution, and stellar-mass black holes \citep{BreenHeggie2013}.

In the right panel of Fig.~\ref{Fig_g_ra} we show the evolution of the ratio of the anisotropy radius to the half-mass radius, $\ra/\rh$, as we did in Table~\ref{Tab_BF_Param}. The variation of the values of this ratio captures well the two phases of the evolution described in Section~\ref{Sect_Results_1_anis}: in the pre-collapse phase, the values of $\ra/\rh$ decrease (i.e., the portion of the cluster characterised by radial anisotropy increases), and after core collapse it progressively increases in time (i.e. the cluster becomes more isotropic).

By inspecting Table~\ref{Tab_BF_Param}, it is clear that there appears to be a sharp transition in the values of this ratio between $3 \times 10^5$ and $3.5 \times 10^5$ $N$-body times. This happens because at that point the cluster becomes isotropic and, as mentioned in Section~\ref{Sect_Models}, the anisotropy radius needs to be larger than the truncation radius, in order to have an isotropic model. In the figure we do not show the values of  $\ra/\rh$ for the snapshots from 7 to 11, because their best-fit models are isotropic, and the values of this ratio become extremely large with respect to those represented there.

\subsubsection{True properties of the cluster}
We have the unique possibility of comparing the results of the fit with the true properties of the stellar system. Fig.~\ref{Fig_properties} shows the comparison of the values of some relevant quantities derived from the fit to those calculated from the snapshots. Orange points and lines represent the values obtained from the best-fit \textsc{limepy} models, and error bars are always plotted; red pentagons mark the cases for which the radial profiles have been shown in Figs.~\ref{Fig_densityEJ}--\ref{Fig_anis}.

The left panels in the figure show the ratio of the mass (top) and half-mass radius (bottom) obtained as best-fit for the models to the true values listed in Table~\ref{Tab_True_Properties}; for comparison, a ratio of unity is represented by a dashed line in the plots. The best-fit values are within 4\% of the real values for both quantities. Except for the last snapshot, the best-fit half-mass radii are usually smaller than the true values, while for the mass sometimes we obtain values slightly larger than the true ones. It is interesting to note that we formally do not have any ``observational error'' in the data, so the discrepancy observed between the best-fit values and the real ones suggests that the models are not perfect in reproducing the simulations, even though they offer a good representation of their principal properties.

Some other quantities, that are not fitting parameters but can be calculated from the models, can be compared to their true values calculated from the snapshots. The top right panel of Fig.~\ref{Fig_properties} shows the evolution in time of the values of $\kappa$ (for a definition, see equation~\ref{Eq_kappa}). The black dots are the true values, the orange and red dots are the ones calculated for the best-fit models. We recall that for the models considered here the anisotropy radius $\ra$ is monotonically related to the parameter $\kappa$, that indicates the amount of anisotropy present in the system: to smaller values of $\ra$ correspond larger values of $\kappa$, and the system is more anisotropic. By inspecting the figure, it is clear that the models overestimate the anisotropy content of the snapshots, as already noted when discussing the anisotropy profiles of Fig.~\ref{Fig_anis}. The value of $\kappa$ calculated from the simulation is smaller than 1 for two snapshots towards the end of the simulation: the models do not allow for the presence of tangentially biased anisotropy, and therefore are incapable of reproducing these values. We note that, however, the values of $\kappa<1$ are probably due to the noise around isotropy, due to the relative low number of particles left in the simulation, that is also observed in the anisotropy profiles of Fig.~\ref{Fig_anis}.

The bottom right panel of Fig.~\ref{Fig_properties} shows the ratio of the truncation radius $\rt$ calculated from the model to the true value of the Jacobi radius for each snapshot. In the pre-collapse phase, initially the models underpredict the values of the tidal radius, then they increasingly overpredict it: just before core collapse, the tidal radius calculated for the best-fit model is more than twice larger than the Jacobi radius of the cluster at that time. After core collapse, except for the first snapshot, for which we find a value of $\rt$ that is compatible with the tidal radius of the cluster, we always obtain a value of the truncation radius roughly corresponding to 80\% of the tidal radius. We note that this could be related to the fact that the outermost bound star in each snapshot is usually located at around 80\% of the tidal radius \citep{Kuepper2010}.

\begin{table}
\begin{center}
\caption[Best-fit King model parameters.]{Best-fit parameters for King models. For each snapshot, identified by a label in the first column, we list the values of the best-fit parameters. Columns from 2 to 5, respectively, are as follows: the concentration $W_0$, the mass of the cluster $M$, the half-mass radius $\rh$, the logarithm of the total mass of stars in the background $\log M_{\rm bg}$.} 
\label{Tab_BF_Param_King}
\begin{tabular}{ccccc}
\hline\hline
Snapshot  & $W_0$ & $M$ & $\rh$ & $\log M_{\rm bg}$ \\
\hline
pre-CC  1 & 7.00 & 1.140 & 1.026 & -2.21 \\
pre-CC  2 & 7.32 & 1.139 & 1.030 & -2.03 \\
pre-CC  3 & 7.55 & 1.157 & 1.084 & -1.93 \\
pre-CC  4 & 7.75 & 1.166 & 1.093 & -1.79 \\
pre-CC  5 & 7.98 & 1.209 & 1.182 & -1.71 \\
pre-CC  6 & 8.23 & 1.235 & 1.260 & -1.63 \\
pre-CC  7 & 8.37 & 1.246 & 1.301 & -1.54 \\
pre-CC  8 & 8.53 & 1.261 & 1.363 & -1.49 \\
pre-CC  9 & 8.74 & 1.260 & 1.408 & -1.42 \\
pre-CC 10 & 9.00 & 1.331 & 1.589 & -1.38 \\
\hline
 1 &  9.34 & 0.953 & 5.087 & -2.07 \\
 2 &  9.54 & 0.667 & 5.573 & -2.78 \\
 3 & 10.03 & 0.498 & 6.059 & -3.22 \\
 4 &  9.86 & 0.390 & 5.893 & -3.82 \\
 5 & 10.08 & 0.313 & 5.964 & -4.59 \\
 6 & 11.57 & 0.251 & 6.230 & -4.59 \\
 7 & 10.29 & 0.182 & 5.634 & -6.82 \\
 8 &  9.81 & 0.133 & 4.875 & -6.78 \\
 9 & 10.79 & 0.087 & 4.628 & -6.83 \\
10 & 10.65 & 0.047 & 3.774 & -6.81 \\
11 &  8.32 & 0.014 & 2.275 & -6.68 \\
\hline
\end{tabular}
\end{center}
\end{table}

\subsection{Comparison with King models}
\label{Sect_Results_3}

The family of \textsc{limepy} models introduced in Section~\ref{Sect_Models} turns out to be a good choice to represent the cluster in the different phases of its evolution. The flexibility given by the truncation parameter, with respect to, for example, the most commonly used \cite{King1966} models, allows us to reproduce in a reasonable way the main properties of the system, especially near the truncation radius. 

For a more detailed comparison, we carried out fits with isotropic King models (by using \textsc{limepy} with $g = 1$, and $\ra = \infty$), and we list the best-fit parameters in Table~\ref{Tab_BF_Param_King}. We also indicate the best-fit profiles for King models with dashed lines in Figs.~\ref{Fig_densityEJ}--\ref{Fig_anis}.

\begin{figure}
\includegraphics[width=0.49\textwidth]{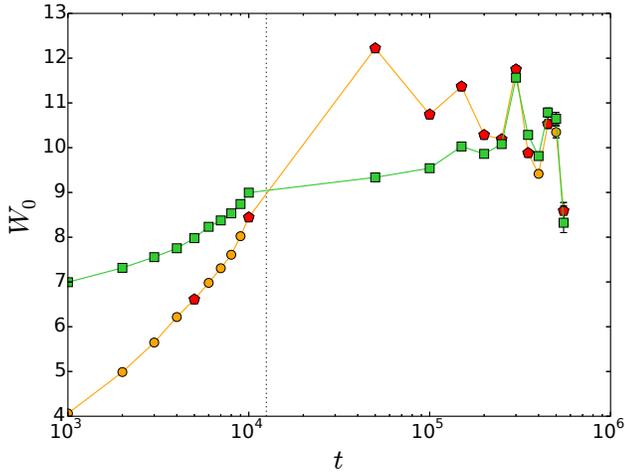}
\caption{Time evolution of the best-fit concentration parameter $W_0$. Orange dots represent the best-fit values obtained from the fit to each snapshot by means of \textsc{limepy} models, and green squares those obtained when considering isotropic King models. A line connecting the points has been added to better show the trend. As done in the previous figures, in the case of \textsc{limepy} models we mark with red pentagons the values corresponding to the snapshots whose relevant profiles are reproduced in Figs.~\ref{Fig_densityEJ}--\ref{Fig_anis}. The vertical dotted line indicates the moment of core collapse.}
\label{Fig_W0}
\end{figure}

King models generally provide a worse fit with respect to anisotropic \textsc{limepy} models, especially for the early snapshots. In particular, the largest discrepancies with respect to the profiles calculated from the simulations are observed at small and large radii. King models generally underpredict the density and the velocity dispersion in the centre; in the outermost parts of the cluster, they underpredict the density, and they overpredict the velocity dispersion. Also, they usually have a smaller truncation radius with respect to the Jacobi radius of the cluster. These shortcomings of the King models may be easily interpreted in light of the assessment in phase space conducted in Sect.~\ref{Sect_Results_0}. In fact, especially in the proximity of the core collapse (e.g., see panels corresponding to Snapshots pre-CC10, 1, 2 of Fig.~\ref{Fig_densityEJ}), when radial anisotropy is the strongest (see Fig.~\ref{Fig_properties} top-right panel), unsurprisingly, isotropic equilibria fail to describe the interplay between energy and angular momentum, particularly at low values of $J^2$. Such a discrepancy affects both the central and the outer parts of the cluster, since, at low energies, best-fit King models tend to favour values of $J^2$ which are too high (i.e., dashed contours are much higher than the red ones), while, at higher energies the role of angular momentum is missed altogether (i.e., the dashed contours stops too early). This behaviour in phase space has immediate impact on the slopes of the velocity moments, both at small and large radii.

As expected, the best representation of the cluster is obtained, with these models, for the late snapshots: we note that the \textsc{limepy} models have best-fit truncation parameters $g$ close to 1 at these times, and the cluster is mainly characterised by isotropy. We notice that King models have a limited range of $\rj/\rh \lesssim 7.5$, hence they will never be able to describe clusters that are deeply embedded in the tidal field \citep{Baumgardt2010}, as it happens in the first stages of the evolution of the cluster analysed here. 

Figure~\ref{Fig_W0} shows the time evolution of the best-fit values of the concentration parameter $W_0$ for \textsc{limepy} and King models, represented with orange (red) and green points, respectively. In both cases, the values of this parameter are increasing in the pre-collapse phase, and for King models the values are always larger than for \textsc{limepy} models. After core collapse, the values obtained for \textsc{limepy} models are decreasing, with some scatter, while those for King models are initially roughly constant and then, in correspondence of the last snapshots, show an oscillatory behaviour, with a range of values which is comparable to the one obtained for \textsc{limepy} models. On this note, we wish to emphasise that, even if the numerical values are comparable, the scale of central concentration traced by the $W_0$ parameter in the case of \cite{King1966} models is different from the one associated with the \textsc{limepy} models, especially in view of the role played by the truncation parameter $g$. We also point out that the values of $W_0$ obtained for the King models in the pre-collapse phase are larger than those obtained in previous studies \citep[see for example][]{Chernoff1986}: this apparent discrepancy is most likely due to the very underfilled initial state of the $N$-body model. 

\begin{figure}
\includegraphics[width=0.49\textwidth]{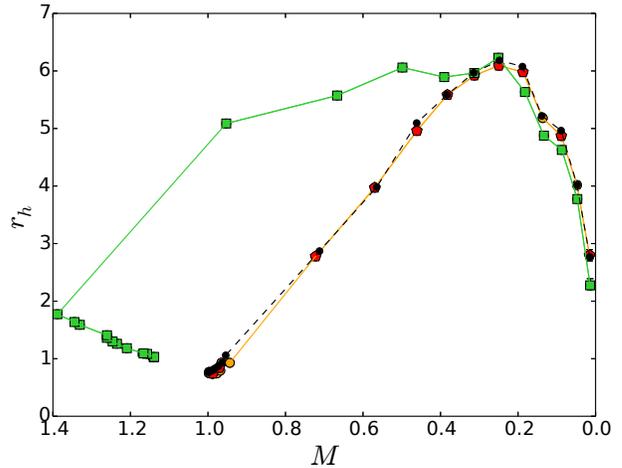}
\caption{Evolution of the values of the mass $M$ and half-mass radius $\rh$ of the cluster: in time, the evolution goes from left to right. Black dots, connected by the black dashed line, represent the true values obtained from the snapshots. Orange dots represent the best-fit values obtained from the fit by means of \textsc{limepy} models, and green squares those obtained when considering isotropic King models. As done in the previous figures, in the case of \textsc{limepy} models we mark with red pentagons the values corresponding to the snapshots whose relevant profiles are reproduced in Figs.~\ref{Fig_densityEJ}--\ref{Fig_anis}. We note that the apparently unphysical increase of the total mass in the pre-collapse phase, as traced by the evolution of the values resulting from the fit with King models, is a particularly crucial shortcoming of such interpretative framework.}
\label{Fig_comparison_King}
\end{figure}

Figure~\ref{Fig_comparison_King} shows the evolution of the mass and of the half-mass radius of the cluster, as compared to the values obtained from the King and \textsc{limepy} models. In the figure, the evolution goes from left to right (decreasing mass in time). We note that King models usually overestimate the mass of the cluster. They also initially overestimate and then underestimate the half-mass radius. The discrepancy is larger for the first part of the evolution, as clearly shown in the figure, especially in the pre-collapse phase, where the best-fit mass increases with time. \cite{GielesHeggieZhao2011} estimated that roughly 2/3 of the Milky Way globular clusters are still in the early expansion phase of their evolution. Our results suggests therefore that isotropic King models are applicable to only 1/3 of the Galactic globular cluster population.

It is remarkable that the simple \textsc{limepy} models used here are able to reproduce the key properties of the snapshots at all stages, in a much more satisfactory way than is possible with King models. The opportunity to describe the snapshots selected in this study by means of a single family of dynamical models with a variable degree of anisotropy is particularly convenient, especially because it allows us to characterise the entire evolution of the stellar system within a single, well-posed dynamical framework. In addition, since the family of \textsc{limepy} models smoothly converges to the family of \cite{King1966} models in the limit of the absence of anisotropy, the comparison between these two frameworks is correctly-set and particularly insightful.


\section{Conclusions}
\label{Sect_Conclusion}

We propose a first application of a recently proposed family of models \citep[the \textsc{limepy} models,][]{GielesZocchi2015} to fit several snapshots of a direct $N$-body simulation, spanning the entire life of a star cluster, with the aim of testing the applicability of simple models to describe the dynamics of star clusters.

The \textsc{limepy} models include a parameter, $g$, that determines the sharpness of the spatial truncation, and another parameter, $\ra$, that regulates the presence of radially biased pressure anisotropy. The flexibility of these models allows us to study different phases in the life of a globular cluster, and to explore the role of tides and anisotropy in time. Indeed, by looking at the evolution of the best-fit values of these parameters we obtain information about the dynamical evolution of the cluster. The parameter $g$ decreases in time, indicating that the truncation becomes steeper and more abrupt due to the effect of the external tidal field. The anisotropy radius $\ra$ decreases in the pre-collapse phase, and then increases in time, showing how a cluster that in the early phases of its evolution developed radial anisotropy later evolves towards an isotropic configuration.

The models appear to be adequate in reproducing the radial profiles of the main observables, such as the density, the velocity dispersion, and the anisotropy profiles, when compared to the ones calculated by binning the stars in the snapshots. The evolution of anisotropy is well reproduced, in time, by these models, even if in the earliest snapshots the model profiles overpredict the ones obtained by binning stars in the snapshots. 

The best-fit values obtained for the mass and the half-mass radius of the cluster are in satisfactory agreement with the true values characterising the snapshots. The truncation radius of the models does not accurately reproduce the position of the tidal radius of the cluster in the pre-collapse phase. After core collapse, it usually has a value that corresponds to 80\% of the tidal radius of the cluster. The possibility to compare the results of the fits to the true values of some relevant quantities is particularly important to determine how well these models can reproduce these stellar systems, and to highlight the possible presence of systematic biases, which should be properly taken into account when the distribution function based models are applied to the interpretation of observational data of Galactic globular clusters.

The results of this investigation are also useful to put limits on the amount of radial anisotropy that can be expected for globular clusters evolving in a tidal field. This is particularly important for studies of line-of-sight velocity dispersion profiles of Galactic globular clusters. The presence of radially-biased anisotropy in the outer parts of these stellar systems causes the central projected velocity dispersion to be larger than the corresponding isotropic case. A similar effect could be due to other factors, such as for example the presence of an intermediate-mass black hole \citep[see for example][and references therein]{ZGHB2015}. Determining the degree of anisotropy present in a cluster is therefore important to determine if an intermediate-mass black hole is present in its centre, and to quantify its mass.

Here we used a discrete fitting technique to fit models to the snapshots, in order to use all the information provided by the data, without degrading it by binning stars to create radial profiles of the quantities under consideration. This technique is promising, and we plan to extend it to fit on real data. To do this, we will need to take into account the fact that observed globular clusters appear projected on the plane of the sky, and each measurement comes with an associated error that should have the proper treatment in a fit. This aspect is crucial, especially in consideration of the forthcoming astrometric information which will be provided for many Galactic globular clusters by the mission Gaia; in this respect, the subsequent step in our programme will be to include in our framework a treatment of the measurement errors which will allow us to make the most of the upcoming era of ``precision astrometry''.

\section*{Acknowledgements}
We thank Antonio Sollima for comments on an earlier version of the manuscript, and the anonymous referee for constructive feedback. AZ acknowledges financial support from the Royal Society (Newton International Fellowship). MG acknowledges the European Research Council (ERC-StG-335936) and the Royal Society for financial support. ALV is grateful to Douglas Heggie for thought-provoking conversations, and to the Royal Commission for the Exhibition of 1851 for financial support. This project was initiated during the Gaia Challenge (\url{http://astrowiki.ph.surrey.ac.uk/dokuwiki}) meeting in 2013 (University of Surrey) and further developed in the follow-up meeting in 2014 (MPIA in Heidelberg). 

\bibliographystyle{mn2e}
\bibliography{biblio.bib}

\appendix
\section{A more quantitative comparison of radial profiles}
\label{App_residuals}

We provide here a more quantitative comparison between the radial profiles calculated from the snapshots and the ones predicted by the best-fit models, in addition to the discussion presented in Sect.~\ref{Sect_Results_1}. To do this, we introduce the following quantity:
\begin{equation}
\mu_x = \dfrac{x_{\rm mod}(r_i) - x_i}{\delta x_i} \ ,
\label{Eq_mu}
\end{equation}
where $x_{\rm mod}$ represents one of the quantities predicted by the models, and $x_i$ is the corresponding value obtained, at the radial position $r_i$ and with an error $\delta x_i$, when binning the stars in the snapshots of our numerical simulation. We notice that $\mu_x$ is the ratio of the residuals to the error of a given quantity $x$. This choice is motivated by the fact that we want to be able to compare the results obtained for different profiles in a similar way. Based on its definition, positive values of $\mu_x$ indicate that the best-fit model overestimates the data, negative values that it underestimates the data.

\begin{figure*}
\includegraphics[width=\textwidth]{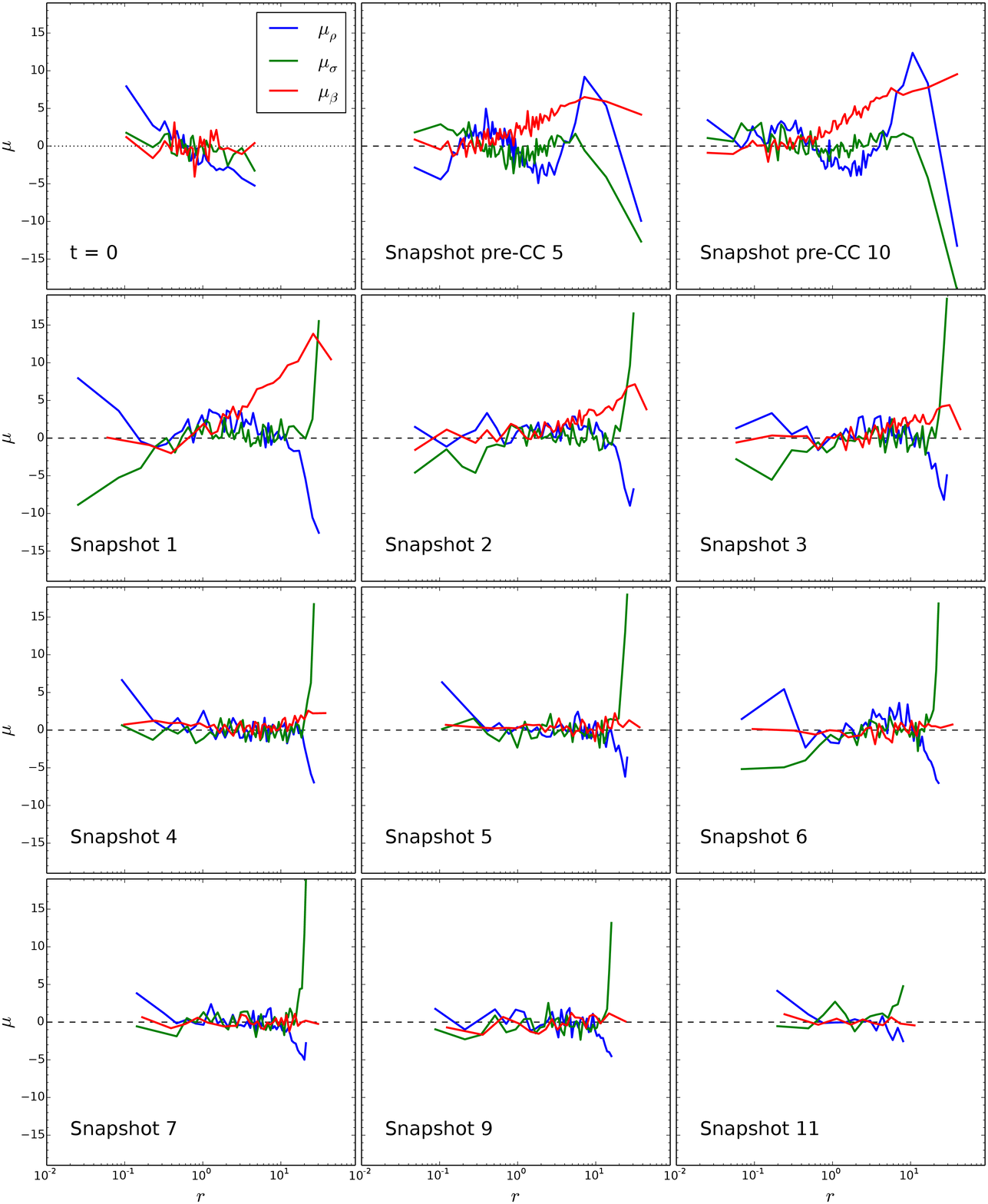}
\caption{Values of the quantities $\mu_{\rho}$, $\mu_{\sigma}$, and $\mu_{\beta}$, calculated according to equation~(\ref{Eq_mu}) for the best-fit \textsc{limepy} models for the density $\rho$, the velocity dispersion $\sigma$, and the anisotropy $\beta$, respectively. Each panel corresponds to a given snapshot, with labels listed in Table~\ref{Tab_True_Properties}. Blue lines show the behaviour of $\mu_{\rho}$, green lines of $\mu_{\sigma}$, and red lines of $\mu_{\beta}$.}
\label{Fig_res_limepy}
\end{figure*}

\begin{figure*}
\includegraphics[width=\textwidth]{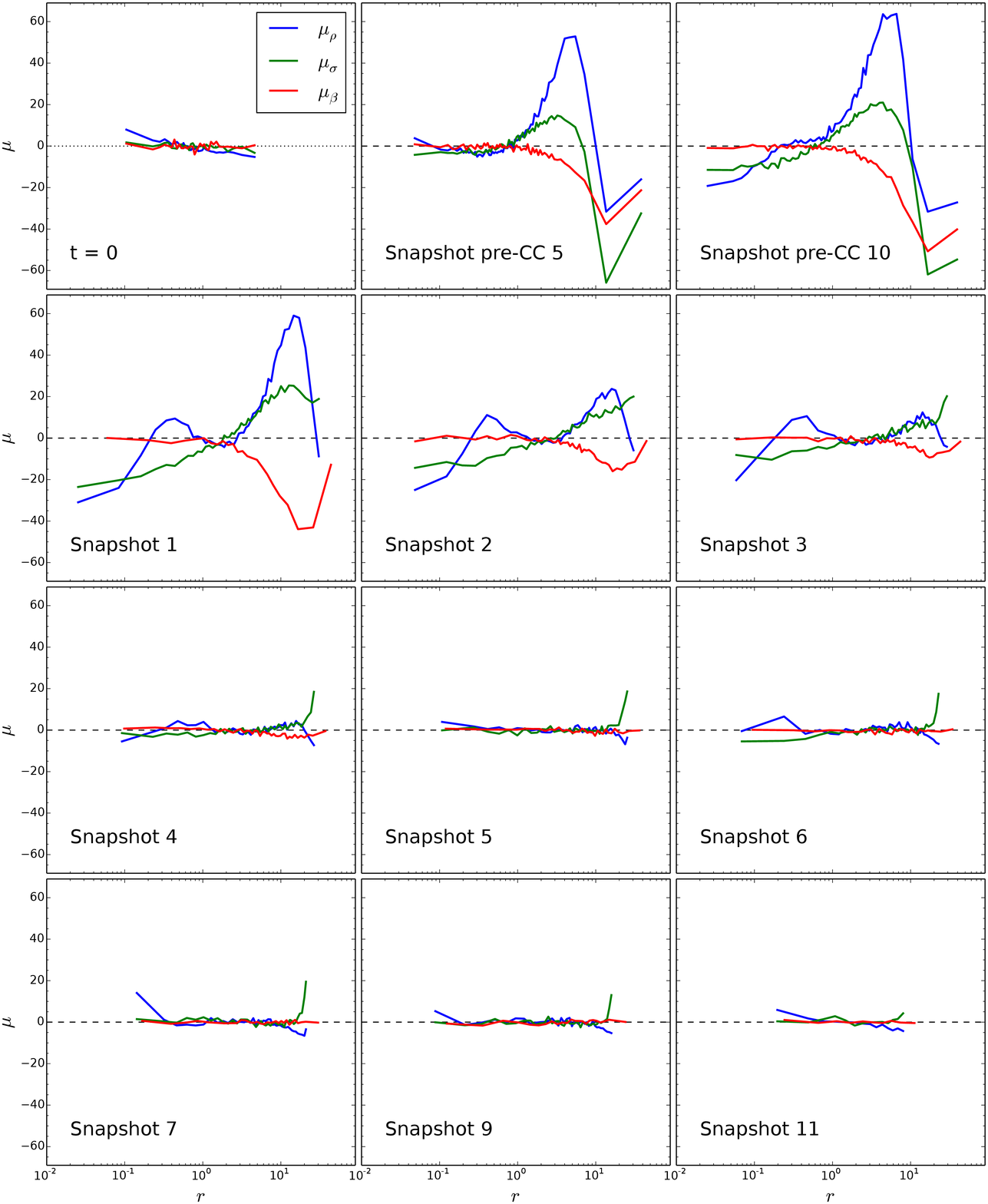}
\caption{Values of the quantities $\mu_{\rho}$, $\mu_{\sigma}$, and $\mu_{\beta}$, calculated according to equation~(\ref{Eq_mu}) for the best-fit King models for the density $\rho$, the velocity dispersion $\sigma$, and the anisotropy $\beta$, respectively. Each panel corresponds to a given snapshot, with labels listed in Table~\ref{Tab_True_Properties}. Blue lines show the behaviour of $\mu_{\rho}$, green lines of $\mu_{\sigma}$, and red lines of $\mu_{\beta}$.}
\label{Fig_res_king}
\end{figure*}

We compute $\mu_{\rho}$, $\mu_{\sigma}$, and $\mu_{\beta}$, by considering the density $\rho$, the velocity dispersion $\sigma$, and the anisotropy $\beta$, respectively. In Fig.~\ref{Fig_res_limepy} we show the values of these functions for the snapshots considered in Figs.\ref{Fig_densityEJ}-\ref{Fig_anis} in the case of \textsc{limepy} models; Fig.~\ref{Fig_res_king} refers instead to King models. Blue lines show the behaviour of $\mu_{\rho}$, green lines of $\mu_{\sigma}$, and red lines of $\mu_{\beta}$. In the top left panel of each figure we show the comparison between the profiles we calculated from the sampled Plummer initial conditions with respect to the corresponding analytical profiles: the deviations here are due to the discrete nature of the initial conditions, and to the binning choice.

By inspecting the figures, it is clear that the largest deviations of the profiles calculated from the snapshots with respect to the best-fit ones are found at early times, particularly before core collapse. The distance between the models and the data is larger when considering King models, as expected based on the discussion of Sect.~\ref{Sect_Results_3}. The largest differences are observed, for each snapshot, in the innermost and outermost radial regions. We notice that for each snapshot and for a given family of models $\mu$ assumes comparable values for the three quantities considered.

\end{document}